\documentclass[12pt,letterpaper]{JHEP3}
\usepackage{cite}
\usepackage{epsfig}
\usepackage{amsmath}
\usepackage{amssymb}
\usepackage{subfigure}

\newcommand{\Ov}[1]{\frac{1}{#1}}
\newcommand{\tobs}{t_{\rm obs}}
\newcommand{\tl}{t_\Lambda}
\newcommand{\tc}{t_{\rm c}}
\newcommand{\al}{\alpha}
\newcommand{\eps}{p}
\newcommand{\mcp}{M_{\rm CP}}
\newcommand{\mah}{M_{\rm AH}}
\newcommand{\tf}{t_{\rm f}}
\newcommand{\tlmax}{\tl^{\rm max}}

\title{Geometric origin of coincidences and hierarchies in the landscape}

\author{Raphael Bousso$^{a,b}$, Ben Freivogel$^{c}$, Stefan Leichenauer$^{a,b}$ and Vladimir Rosenhaus$^{a,b}$\\ \\
  $^a$ Center for Theoretical Physics and Department of Physics\\
\ \  University of California, Berkeley, CA 94720-7300, U.S.A.\\
$^b$  Lawrence Berkeley National Laboratory, Berkeley, CA 94720-8162,
  U.S.A.\\
$^c$\! Center for Theoretical Physics and Laboratory for Nuclear Science\\
\ \ Massachusetts Institute of Technology, Cambridge, MA 02139, U.S.A.}

\abstract{We show that the geometry of cutoffs on eternal inflation strongly constrains predictions for the timescales of vacuum domination, curvature domination, and observation.  We consider three measure proposals: the causal patch, the fat geodesic, and the apparent horizon cutoff, which is introduced here for the first time.  We impose neither anthropic requirements nor restrictions on landscape vacua.  For vacua with positive cosmological constant, all three measures predict the double coincidence that most observers live at the onset of vacuum domination and just before the onset of curvature domination.  The hierarchy between the Planck scale and the cosmological constant is related to the number of vacua in the landscape.  These results require only mild assumptions about the distribution of vacua (somewhat stronger assumptions are required by the fat geodesic measure).  At this level of generality, none of the three measures are successful for vacua with negative cosmological constant.  Their applicability in this regime is ruled out unless much stronger anthropic requirements are imposed.}

 \preprint{}
\begin{document}

\section{Introduction}
\label{sec-intro}

String theory appears to contain an enormous landscape of metastable
vacua~\cite{BP,KKLT}, with a corresponding diversity of low-energy
physics.  The cosmological dynamics of this theory is eternal
inflation. It generates a multiverse in which each vacuum is
produced infinitely many times.

In a theory that predicts a large universe, it is natural to assume
that the  relative probability for two different
outcomes of an experiment is the ratio of the expected
number of times each outcome occurs.   But in eternal inflation,
every possible outcome happens infinitely many times.  The relative
abundance of two different observations is ambiguous until one defines
a measure: a prescription for regulating the infinities of eternal
inflation.

Weinberg's prediction~\cite{Wei87} of the cosmological
constant~\cite{Rie98,Per98} was a stunning success for this type of
reasoning.  In hindsight, however, it was based on a measure that was
ill-suited for a landscape in which parameters other than $\Lambda$
can vary.  Moreover, the measure had severe phenomenological
problems~\cite{BouFre06b}.  This spurred the development of more
powerful measure proposals in recent
years
~\cite{LinMez93,LinLin94, GarLin94,GarLin94a,GarLin95,
  GarSch05,VanVil06,Van06,Bou06,Lin06,Lin07,Pag08,
  GarVil08,Win08a,Win08b,Win08c,LinVan08,Bou09, BouFre10, SchVil10,Pag10}.  Surprisingly, some of these
measures do far more than to resolve the above shortcomings.  As we
shall see in this paper, they obviate the need for Weinberg's
assumption that observers require galaxies; and they help overcome the
limitation of fixing all parameters but one to their observed values.

In this paper we will analyze three different measure proposals.  Each regulates the infinite multiverse by restricting attention to a finite portion.  The causal patch measure~\cite{Bou06} keeps the causal past of the future endpoint of a geodesic; it is equivalent to a global cutoff known as light-cone time~\cite{Bou09,BouYan09}.  The fat geodesic measure~\cite{BouFre08b} keeps a fixed physical volume surrounding the geodesic; in simple situations, it is equivalent to the global scale factor time cutoff~\cite{DGSV08}. We also introduce a new measure, which restricts to the interior of the apparent horizon surrounding the geodesic.  

From little more than the {\em geometry\/} of these cutoffs, we are
able to make remarkable progress in addressing cosmological
coincidence and hierarchy problems.  Using each measure, we will
predict three time scales: the time when observations are made,
$\tobs$, the time of vacuum energy domination, $\tl \equiv
\sqrt{3/|\Lambda|}$, and the time of curvature domination,
$\tc$.\footnote{In regions where curvature will never dominate, such
  as our own vacuum, $\tc$ is defined as the time when curvature would
  come to dominate if there were no vacuum energy.  Since our
  observations are consistent with a flat universe, we can only place
  a lower bound on the observed value of $t_{\rm c}$.  We include
  $t_{\rm c}$ in our analysis because bubble universes in the
  multiverse are naturally very highly curved, so the absence of
  curvature requires an explanation.  Moreover, we shall see that some
  measures select for high curvature in vacua with with negative
  cosmological constant. This effect is partly responsible for the
  problems encountered in this portion of the landscape.} We work in
the approximation that observations occur in nearly homogeneous FRW
universes so that these time scales are well defined.

We will allow all vacuum parameters to vary simultaneously.  Parameters for which we do not compute probabilities are marginalized.  We will not restrict attention to vacua with specific features such as baryons or stars.  We will make some weak, qualitative assumptions about the prior distribution of parameters in the landscape.  We will assume that most observers are made of something that redshifts faster than curvature.  (This includes all forms of radiation and matter but excludes, e.g., networks of domain walls.)  But we will not impose any detailed anthropic requirements, such as the assumption that observers require galaxies or complex molecules; we will not even assume that they are correlated with entropy production~\cite{Bou06,BouHar07,BouHar10}.  Thus we obtain robust predictions that apply to essentially arbitrary observers in arbitrary vacua.

The probability distribution over all three variables can be decomposed into three factors, as we will explain further in Sec.~\ref{sec-counting}:
\begin{align} 
\lefteqn{\frac{d^3p}{d\log \tobs ~d \log \tl~ d\log \tc} =}\nonumber \\
& & \frac{d^2 \tilde p}{~d\log \tl~d\log \tc} \times M(\log \tobs, \log \tc,
\log \tl) \times \al(\log \tobs,\log  \tc,\log \tl)~.
\end{align}
Here $ \tilde p $ is the probability density that a bubble with parameters $(\log \tl, \log \tc)$ is produced within the region defined by the cutoff.  Its form can be estimated reliably enough for our purposes from our existing knowledge about the landscape.  The factor $M(\log \tobs, \log \tl, \log \tc)$ is the mass inside the cutoff region at the time $\tobs$. This is completely determined by the geometry of the cutoff and the geometry of an FRW bubble with parameters $(\log t_\Lambda, \log t_{\rm c})$, so it can be computed unambiguously.  The last factor, $\al(\log \tobs,  \log \tc, \log \tl)$, is the one we know the least about. It is the number of observations per unit mass per logarithmic time interval, averaged over all bubbles with the given values $(\log \tl,\log \tc)$.

A central insight exploited in this paper is the following.  In bubbles with positive cosmological constant, the calculable quantity $M$ so strongly suppresses the probability in other regimes that in many cases we only need to know the form of $\al$ in the regime where observers live before vacuum energy or curvature become important, $\tobs \lesssim \tl, \tc$.  Under very weak assumptions, $\al$ must be independent of $t_\Lambda $ and $t_{\rm c} $ in this regime.  This is because neither curvature nor vacuum energy play a dynamical role before observers form, so that neither can affect the number of observers per unit mass.  Thus, for positive cosmological constant $\alpha$ is a function only of $t_{\rm obs}$ in the only regime of interest.  The success of the measures in explaining the hierarchy and coincidence of the three timescales depends on the form of this function.  We will that the causal patch and apparent horizon cutoff succeed well in predicting the three timescales already under very weak assumptions on $\alpha$.  The fat geodesic cutoff requires somewhat stronger assumptions.
 
For negative cosmological constant, however, the geometric factor $M$ favors the regime where $\tobs$ is not the shortest time scale.  Thus, predictions depend crucially on understanding the form of $\alpha$ in this more complicated regime.  For example, we need to know on average to what extent early curvature domination disrupts the formation of observers. What is clear from our analysis is that all three measures are in grave danger of predicting that most observers see negative $\Lambda$, in conflict with observation, unless the anthropic factor $\alpha$ takes on rather specific forms.  Assuming that this is not the case, we must ask whether the measures can be modified to agree with observation. Both the causal patch measure and the fat geodesic measure are dual~\cite{BouFre08b,BouYan09} to global time cutoffs, the lightcone time and the scale factor time cutoff, respectively. These global cutoffs, in turn, are motivated by an analogy with the UV/IR relation in AdS/CFT~\cite{GarVil08}. But this analogy really only applies to positive $\Lambda$, so it is natural to suspect that the measure obtained from it is inapplicable to regions with $\Lambda\leq 0$~\cite{Bou09}. (Indeed, the causal patch does not eliminate infinities in $\Lambda=0$ bubbles~\cite{Bou06}.  We do not consider such regions in this paper.)
 
\paragraph{Outline and summary of results} 

In Sec.~\ref{sec-counting}, we will describe in detail our method for counting observations.  We will derive an equation for the probability distribution over the three variables $(\log \tl, \log \tc, \log\tobs)$.  We will explain how simple qualitative assumptions about $\alpha(\log \tobs,  \log \tc, \log \tl)$, the number of observations per unit mass per unit logarithmic time interval, allow us to compute probabilities very generally for all measures.

We work in the approximation that observations in the multiverse take
place in negatively curved Friedmann-Robertson-Walker (FRW)
universes. In Sec.~\ref{sec-frw}, we will obtain solutions for their
scale factor, in the approximation where the matter-, vacuum-, and
possibly the curvature-dominated regime are widely separated in time.

In Secs.~\ref{sec-cp}--\ref{sec-sf}, we will compute the probability
distribution over $(\log \tl, \log \tc, \log\tobs)$, using three
different measures.  For each measure we consider separately the cases
of positive and negative cosmological constant. As described above,
the results for negative cosmological constant are problematic for all
three measures.  We now summarize our results for the case $\Lambda > 0$.

In Sec.~\ref{sec-cp}, we show that the causal patch measure predicts the double coincidence $\log t_{\rm c} \approx \log \tl \approx \log \tobs$.  We find that the scale of all three parameters is related to the number of vacua in the landscape.  This result is compatible with current estimates of the number of metastable string vacua.  Such estimates are not sufficiently reliable to put our prediction to a conclusive test, but it is intriguing that the size of the landscape may be the origin of the hierarchies we observe in Nature (see also~\cite{Pol06,Bou06,Bou07,BouHal09,BouLei09,BouHar10}.    We have previously reported the result for this subcase in more detail~\cite{BouFre10c}.   

Unlike the causal patch, the new ``apparent horizon measure'' (Sec.~\ref{sec-ah}) predicts the double coincidence $\log t_{\rm c} \approx \log \tl \approx \log \tobs$ for any fixed value $t_{\rm obs}$.  When all parameters are allowed to scan, its predictions agree with those of the causal patch, with mild assumptions about the function $\alpha$.  The apparent horizon measure is significantly more involved than the causal patch: it depends on a whole geodesic, not just on its endpoint, and on metric information, rather than only causal structure. If our assumptions about $\alpha$ are correct, this measure offers no phenomenological advantage over its simpler cousin and may not be worthy of further study.

The fat geodesic cutoff, like the apparent horizon cutoff, predicts the double coincidence $\log t_{\rm c} \approx \log \tl \approx \log \tobs$ for any fixed value of $t_{\rm obs}$.  However, it favors small values of all three timescales unless either (a) an anthropic cutoff is imposed, or (b) it is assumed that far more observers form at later times, on average, than at early times, in vacua where curvature and vacuum energy are negligible at the time $t_{\rm obs}$.   (Qualitatively, the other two measures also require such an assumption, but quantitatively, a much weaker prior favoring late observers suffices.)  In the latter case (b), the results of the previous two measures are reproduced, with all timescales set by the size of the landscape.  In the former case (a), the fat geodesic would predict that observers should live at the earliest time compatible with the formation of any type of observers (in any kind of vacuum).  It is difficult to see why this minimum value of $\tobs$ would be so large as to bring this prediction into agreement with the observed value, $\tobs \sim 10^{61}$.

\section{Counting Observations}
\label{sec-counting}

In this section, we explain how we will compute the trivariate probability distribution over  $(\log \tobs, \log \tc, \log \tl)$.  We will clarify how we isolate geometric effects, which can be well-computed for each cutoff, from anthropic factors; and we explain why very few assumptions are needed about the anthropic factors once the geometric effects have been taken into account.  

Imagine labelling every observation within the cutoff region by $(\log \tobs, \log \tc, \log \tl)$. We are interested in counting the number of observations as a function of these parameters. It is helpful to do this in two steps. First, we count bubbles, which are labelled by $(\log \tc, \log \tl)$ to get the ``prior" probability distribution
\begin{equation}
\frac{d^2 \tilde p}{~d\log \tl~d\log \tc}~.
\end{equation}
This $\tilde p$ is the probability density to nucleate a bubble with the given values of the parameters inside the cutoff region.

The next step is to count observations within the bubbles. A given bubble of vacuum $i$ will have observations at a variety of FRW times. In the global description of eternal inflation, each bubble is infinite and contains an infinite number of observations if it contains any, but these local measures keep only a finite portion of the global spacetime and hence a finite number of observations. We parameterize the probability density for observations within a given bubble as
\begin{equation}
\frac{dN_i}{d\log \tobs}\sim
M(\log \tobs, \log \tc, \log \tl) \alpha_i(\log \tobs) ~,
\end{equation}
where $M$ is the mass inside the cutoff region, and $\alpha_i$ is the number of observations per unit mass per logarithmic time interval inside a bubble of type $i$. In this decomposition, $M$ contains all of the information about the cutoff procedure.  For a given cutoff, $M$ depends only on the three parameters of interest. Because we are considering geometric cutoffs, the amount of mass that is retained inside the cutoff region does not depend on any other details of the vacuum $i$. On the other hand, $\alpha_i$ depends on details of the vacuum, such as whether observers form, and when they form; but it is independent of the cutoff procedure.

Since we are interested in analyzing the probability distribution over three variables, we now want to average $\alpha_i$ over the bubbles with a given $(\log \tl,\log \tc)$, to get the average number of observations per unit mass per logarithmic time $\alpha$.  With this decomposition, the full probability distribution over all three variables is
\begin{equation}
\frac{d^3p}{d\log \tobs ~d \log \tl~ d\log \tc} =
\frac{d^2 \tilde p}{~d\log \tl~d\log \tc} M(\log \tobs, \log \tc,
\log \tl) \al(\log \tobs,\log  \tc,\log \tl)~.
\end{equation}
To recap, $ \tilde p $ is the probability for the formation of a bubble with parameters $(\log \tl, \log \tc)$ inside the cutoff region; $M(\log \tobs, \log \tl, \log \tc)$ is the mass inside the cutoff region at the FRW time $\tobs$ in a bubble with parameters $(\log \tl, \log \tc)$; and $\al(\log \tobs,  \log \tc, \log \tl)$ is the number of observations per unit mass per logarithmic time interval, averaged over all bubbles with parameters $(\log \tl,\log \tc)$.

This is a useful decomposition because the mass $M$ inside the cutoff
region can be computed exactly, since it just depends on geometrical
information. We will assume that $d^2 \tilde p/d\log \tl~d\log \tc$ can be factorized into contribution from $\tl$ and a contribution from $\tc$.  Vacua with $\Lambda\sim 1$ can be excluded since they contain only a few bits of information in any causally connected region.   In a large landscape, by Taylor expansion about $\Lambda=0$, the prior for the cosmological constant is flat in $\Lambda$ for $\Lambda\ll 1$, $d\tilde p/d\Lambda =$ const.  Transforming to the variable $\log t_\Lambda$, we thus have  
\begin{equation}
\frac{d^2\tilde p}{d\log \tl d \log \tc} \sim \tl^{-2} g(\log \tc)~.
\end{equation}
The factor $g(\log \tc)$ encodes the prior probability distribution over the time of curvature domination. We will assume that $g$ decreases mildly, like an inverse power of $\log \tc$. (Assuming that slow-roll inflation is the dominant mechanism responsible for the delay of curvature domination, $\log \tc$ corresponds to the number of $e$-foldings. If $g$ decreased more strongly, like an inverse power of $\tc$, then inflationary models would be too rare in the landscape to explain the observed flatness.)  The detailed form of the prior distribution over $\log \tc$ will not be important for our results; any mild suppression of large $\log \tc$ will lead to similar results.

With these reasonable assumptions, the probability distribution becomes
\begin{equation}
\frac{d^3p}{d\log \tobs ~d \log \tl~ d\log \tc} =
\tl^{-2} M(\log \tobs,\log \tc,\log
\tl) g(\log \tc) \al(\log \tobs, \log \tc, \log \tl)~.
\end{equation}
Because $\alpha$ depends on all three variables, it is very difficult to compute in general. However, it will turn out that we can make a lot of progress with a
few simple assumptions about $\alpha$.  First, we assume that in the regime, $\tobs \ll t_\Lambda$, $\alpha$ is independent of $t_\Lambda$.  Similarly, we assume that in the regime, $\tobs \ll t_{\rm c}$, $\alpha$ is independent of $t_{\rm c}$. By these assumptions, in the regime where the observer time is the shortest timescale, $t_{\rm obs} \lesssim t_\Lambda, \tc$, the anthropic factor $\alpha$ will only depend on $\log \tobs$:
\begin{equation}
\al(\log \tobs,\log \tc,\log \tl) \approx \alpha(\log \tobs) \ {\rm for} \ \tobs \lesssim
\tl , \tc ~.
\end{equation}
These assumptions are very weak.  Because curvature and $\Lambda$ are not dynamically important before $\tc$ and $\tl$, respectively, they cannot impact the formation of observers at such early times.  One could imagine a correlation between the number of $e$-foldings and the time of observer formation even in the regime $\tobs \ll \tc$, for example if each one is tied to the supersymmetry breaking scale, but this seems highly contrived.  In the absence of a compelling argument to the contrary, we will make the simplest assumption.

Second, we assume that when either $\tobs \gg \tc$ or $t_{\rm obs}\gg \tl$, $\al$ is not enhanced compared to its value when $\tobs$ is the shortest timescale.  This is simply the statement that early curvature and vacuum domination does not help the formation of observers.  This assumption, too, seems rather weak.  With this in mind, let us for the time being overestimate the number of observers by declaring $\al$ to be completely independent of $\log \tl$ and $\log \tc$:
\begin{equation} 
\al(\log \tobs, \log \tc,\log \tl) \approx \alpha(\log \tobs) ~. 
\end{equation}
This is almost certainly an overestimate of the number of observations in the regime where $\tobs$ is not the shortest time scale. However, we will show that the predictions for positive cosmological constant are insensitive to this simplification because the geometrical factor $M$, which we can compute reliably, suppresses the contribution from this regime. We will return to a more realistic discussion of $\al$ when we are forced to, in analyzing negative cosmological constant.

With these assumptions and approximations,  the three-variable probability distribution takes a more tractable form,
\begin{equation}
{
\frac{d^3p}{d\log \tobs ~d \log \tl~ d\log \tc} = \tl^{-2} g(\log \tc) M(\log \tobs, \log \tl,\log \tc) \alpha(\log \tobs)~. }
\end{equation}
This is the formula that we will analyze in the rest of the paper.  The only quantity that depends on all three variables is the mass, which we can compute reliably for each cutoff using the geometry of open bubble universes, to which we turn next.

\section{Open FRW universes with cosmological constant}
\label{sec-frw}

In this section, we find approximate solutions to the scale factor, for flat or negatively curved vacuum bubbles with positive or negative cosmological constant.  The landscape of string theory contains a large number of vacua that form a ``discretuum'', a dense spectrum of values of the cosmological constant~\cite{BP}.  These vacua are populated by Coleman-DeLuccia bubble nucleation in an eternally inflating spacetime, which produces open Friedmann-Robertson-Walker (FRW) universes~\cite{CDL}.  Hence, we will be interested in the metric of FRW universes with open spatial geometry and nonzero cosmological constant $\Lambda$.   The metric for an open FRW universe is 
\begin{equation}
ds^2 = -dt^2 + a(t)^2 (d\chi^2 +\sinh^2\chi d \Omega_2^2)~.
\end{equation}
The evolution of the scale factor is governed by the Friedmann equation:
\begin{equation} \label{eq:Fried}
\left(\frac{\dot{a}}{a}\right)^2 = \frac{\tc}{a^3} + \frac{1}{a^2} \pm \Ov{t_{\Lambda}^2}~.
\end{equation}
Here $\tl = \sqrt{3/|\Lambda|}$ is the timescale for vacuum domination, and $\tc$ is the timescale for curvature domination.  The term $\rho_m\sim \tc/a^3$ corresponds to the energy density of pressureless matter.  (To consider radiation instead, we would include a term $\rho_{\rm rad}\sim \tc ^2/a^4$; this would not affect any of our results qualitatively.)  The final term is the vacuum energy density, $\rho_\Lambda$; the``$+$'' sign applies when $\Lambda>0$, and the``$-$'' sign when $\Lambda<0$.     

We will now display approximate solutions for the scale factor as a function of FRW time $t$.  There
are four cases, which are differentiated by the sign of $\Lambda$ and
the relative size of $\tc$ and $\tl$. We will compute all
geometric quantities in the limit where the three time scales $t$,
$\tc$, and $\tl$ are well-separated, so that some terms on
the right hand side of Eq.~(\ref{eq:Fried}) can be neglected.  In this
limit we obtain piecewise solution for the scale factor.  We will {\em
  not\/} ensure that these solutions are continuous and differentiable
at the crossover times.  This would clutter our equations, and it
would not affect the probability distributions we compute later. Up to
order-one factors, which we neglect, our formulas are applicable even
in crossover regimes.

If $\tc\gg \tl$, curvature never comes to dominate.  (One can still define $\tc$ geometrically, or as the time when curvature would have dominated in a universe with $\Lambda=0$.) In this limit the metric can be well approximated as that of a perfectly flat FRW universe, and so becomes independent of $\tc$.   We implement this case by dropping the term $\tc/a^3$ in Eq.~(\ref{eq:Fried}).

\paragraph{Positive cosmological constant}
We begin with the case $\Lambda>0$ and $\tc \ll \tl$.  By solving Eq.~(\ref{eq:Fried}) piecewise, we find
\begin{equation} \label{eq:a}
a(t)\sim \left\{\begin{array}{ll}
\tc^{1/3} t^{2/3}~,& t<\tc \\
t~, & \tc<t<\tl \\
\tl e^{t/\tl-1}~, & \tl < t~.
\end{array}\right.
\end{equation}
If $\tc \gg \tl$, there is no era of curvature domination, and the universe can be approximated as flat throughout.  The scale factor takes the form
\begin{equation} \label{eq:aflat}
a(t)\sim \left\{\begin{array}{ll}
\tc^{1/3} t^{2/3}~,& t<\tl \\
\tc^{1/3} \tl^{2/3} e^{t/\tl-1}~, & \tl < t~.
\end{array}\right.
\end{equation}

\paragraph{Negative cosmological constant}
For $\Lambda<0$, the scale factor reaches a maximum and then begins to decrease.  The universe ultimately collapses at a time $t_{\rm f}$, which is of order $\tl$:  
\begin{equation}
t_{\rm f}\approx \pi \tl~.
\end{equation}
The evolution is symmetric about the turnaround time, $t_{\rm f}/2\approx \pi \tl /2$.

Again, we consider the cases $\tl \gg \tc$ and $\tl \ll \tc$ separately.  For $\tc \ll \tl$, the scale factor is
\begin{equation}
a(t)\sim \left\{\begin{array}{ll}
\tc^{1/3} t^{2/3}~,& t<\tc \\
\tl \sin (t/\tl)~, & \tc<t<\tc' \\
\tc^{1/3} (t')^{2/3}~, & \tc'<t~.
\end{array}\right.
\label{eq-negsf}
\end{equation}
We have defined $t'\equiv t_{\rm f}-t$.  

There is no era of curvature domination if $\tc \gtrsim t_{\rm f}/2$.  For $\tc \gg t_{\rm f}/2$, we treat the universe as flat throughout, which yields the scale factor
\begin{equation}\label{eq-negflata}
a(t) \sim t_{\Lambda}^{2/3} \tc^{1/3} \sin^{2/3}(\pi t/t_{\rm f})~,
\end{equation}
where $t_{\rm f}$ here takes on a slightly different value compared to the curved case:
\begin{equation}
t_{\rm f} = 2\pi \tl/3~.
\end{equation}
At a similar level of approximation as we have made above, this solution can be approximated as
\begin{equation}
a(t)\sim \left\{\begin{array}{ll}
\tc^{1/3} t^{2/3}~,& t<t_{\rm f}/2 \\
\tc^{1/3} (t')^{2/3}~, & t_{\rm f}/2<t~.
\end{array}\right.
\label{eq-approxnegflata}
\end{equation}
where again $t' \equiv \tf - t$~.

\section{The causal patch cut-off}
\label{sec-cp}  

With all our tools lined up, we will now consider each measure in turn and derive the probability distribution.  We will treat positive and negative values of the cosmological constant separately.  After computing $M$, we will next calculate the bivariate probability distribution over $\log \tl$ and $\log \tc$, for fixed $\log \tobs$.  In all cases this is a sharply peaked function of the two variables $\log \tl$ and $\log \tc$, so we make little error in neglecting the probability away from the peak region.  Then we will go on to find the full distribution over all three parameters. In this section, we begin with the causal patch measure~\cite{Bou06,BouFre06a}, which restricts to the causal past of a point on the future boundary of spacetime.  The causal patch measure is equivalent~\cite{BouYan09} to the light-cone time cutoff~\cite{Bou09,BouFre10}, so we will not discuss the latter measure separately.

We may use boost symmetries to place the origin of the FRW bubble of interest at the center of the causal patch.  The boundary of the causal patch is given by the past light-cone from the future end point of the comoving geodesic at the origin, $\chi=0$:
\begin{equation} \label{eq:eh}
\chi_{\rm CP}(t) = \int_{t}^{t_{\rm f}}{\frac{dt'}{a(t')}}~.
\end{equation}
If $\Lambda<0$, $t_{\rm f}$ is the time of the big crunch (see Sec.~\ref{sec-frw}).  For long-lived metastable de~Sitter vacua ($\Lambda>0$), the causal patch coincides with the event horizon.  It can be computed as if the de~Sitter vacuum were eternal ($t_{\rm f}\to\infty$), as the correction from late-time decay is negligible.

\subsection{Positive cosmological constant}
\label{sec-cppos}

We begin with the case $\Lambda>0$, $\tc <\tl$.  Using Eq.~(\ref{eq:a}) for $a(t)$, we find
\begin{equation}
\chi_{\rm CP}(\tobs)\sim \left\{\begin{array}{ll}
1+\log(\tl/\tc)+ 3\left[1-(\tobs/\tc)^{1/3}\right]
~,& \tobs<\tc \\
1 +\log (\tl/\tobs)~,& \tc<\tobs<\tl \\
e^{-\tobs/\tl}~,& \tl<\tobs ~.
\end{array}\right.~.
\end{equation}

The comoving volume inside a sphere of radius $\chi$ is $\pi (\sinh 2
\chi_{\rm CP} - 2 \chi_{\rm CP})$.  We approximate this, dropping
constant prefactors that will not be important, by $\chi^3$ for $\chi \lesssim 1$, and by $e^{2\chi}$ for $\chi \gtrsim 1$: 
\begin{equation}
V_{\rm CP} \sim \left\{\begin{array}{ll}
\exp(2 \chi_{\rm CP}) ~,& \tobs<t_{\Lambda} \\
\chi_{\rm CP}^3 ~, & t_{\Lambda}<\tobs ~.
\end{array}\right.~
\end{equation}
The mass inside the causal patch is $M_{\rm CP} = \rho a^3 V_{\rm CP}= \tc V_{\rm CP}$.
\begin{equation} 
\mcp \sim  \left\{\begin{array}{lll}
\tl^2/\tc~, & \tobs<\tc < \tl \ \ & I\\
\tl^2 \tc/ \tobs^2~,& \tc<\tobs<\tl  \ \ & II \\
\tc e^{-3\tobs/\tl}~, & \tc< \tl< \tobs \ \ &III
\end{array}\right.
\label{eq-mcppos}
\end{equation}

Next, we consider the case $\Lambda>0$, $t_{\Lambda}<\tc$.   The above caclulations can be repeated for a flat universe, which is a good approximation for this case:
\begin{equation} 
\mcp \sim  \left\{\begin{array}{lll}
\tl~, & \tobs<t_{\Lambda}   \ \  & V\\
\tl e^{-3\tobs/\tl}~, & \tl<\tobs \ \ & IV
\end{array}\right.
\label{eq-flatpospatchmass}
\end{equation}
The same result could be obtained simply by setting $\tc = t_{\Lambda}$ in (\ref{eq-mcppos}).

The full probability distribution is given by multiplying the mass in
the causal patch by the prior distribution and the number of
observations per unit mass per unit time to give
\begin{equation} 
\frac{d^3p_{\rm CP}}{d\log \tc~d\log \tl d \log \tobs} \sim  g \alpha \times \left\{\begin{array}{lll}
\vspace{.05in}\displaystyle{\frac{1}{\tc}}~, & \tobs<\tc
< \tl \ \ & I\\
\vspace{.05in}
\displaystyle{\tc  \over \tobs^2}~,& \tc<\tobs<\tl  \ \ & II \\
\vspace{.05in}\displaystyle{\tc  \over \tl
  ^2} \exp \left(- \displaystyle{ 3
      \tobs \over \tl} \right) ~, & \tc< \tl< \tobs \ \ &III \\
\vspace{.05in}\displaystyle{1 \over \tl} \exp \left(- \displaystyle{ 3
      \tobs \over \tl} \right)~, & \tl < \tobs , \tc \ \
& IV \\
\displaystyle{1 \over \tl}~, & \tobs < \tl < \tc \ \ & V 
\end{array}\right.
\label{eq-pcppos}   
\end{equation}
Recall that $g(\log \tc)$ is the prior distribution on the time of curvature domination, and $\alpha$ is the number of observations per unit mass per logarithmic time interval.

\begin{figure}[tbp]
\centering
\subfigure[$\Lambda>0$]{
   \includegraphics[width=2.5in]{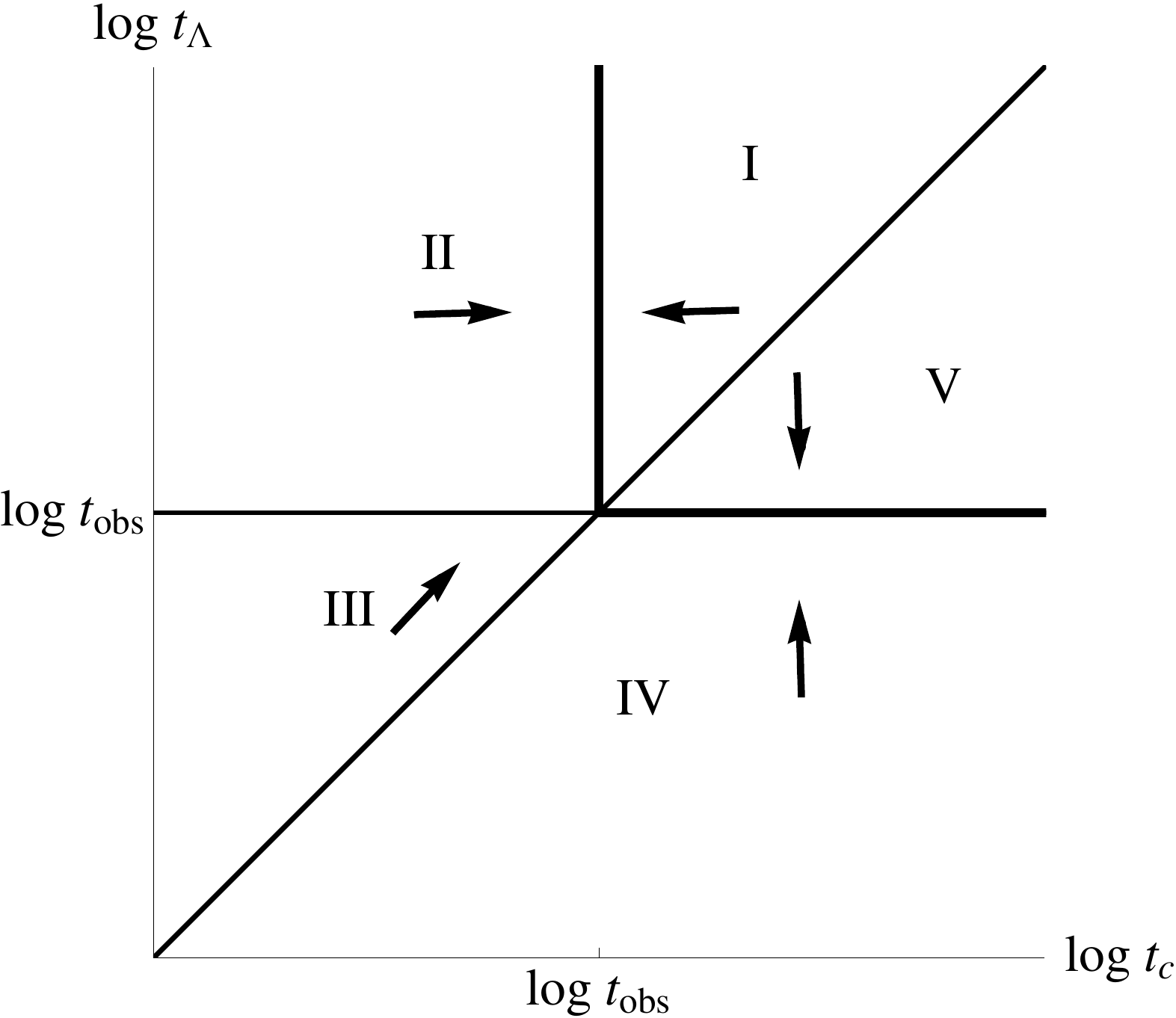}
   }
   \hspace{.5in}
    \subfigure[$\Lambda<0$]{
    \includegraphics[width=2.5 in]{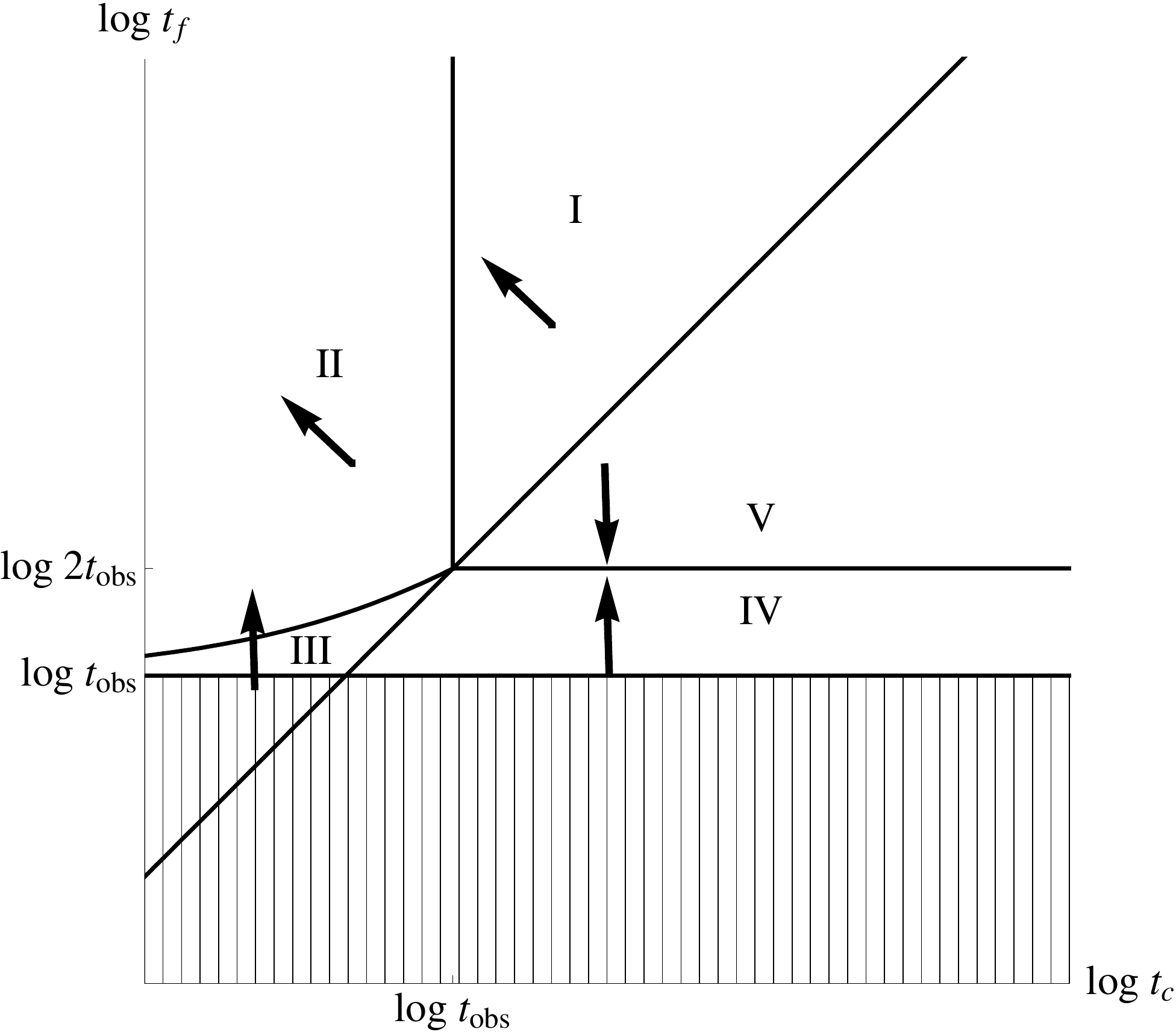}
    }
   \caption{The probability distribution over the timescales of curvature and vacuum domination at fixed observer timescale $\log t_{\rm obs}$, before the prior distribution over $\log t_{\rm c}$ and the finiteness of the landscape are taken into account. The arrows indicate directions of increasing probability.  For $\Lambda>0$ (a), the distribution is peaked along the degenerate half-lines forming the boundary between regions I and II and the boundary between regions IV and V.  For $\Lambda<0$ (b), the probability distribution exhibits a runaway toward the small $\tc$, large $\tl$ regime of region II.   The shaded region is is excluded because $\tobs > t_{\rm f} = \pi \tl$ is unphysical.}
   \label{fig-patchflow} 
\end{figure}

We will first analyze this probability distribution for fixed
$\log \tobs$. As explained in the introduction, we will for the time being overestimate the number of observers in the regime where $\tobs$ is not the shortest timescale by assuming that $\alpha$ is a function of $\log \tobs $ only. We will find that the overestimated observers do not dominate the probability distribution, so this is a good approximation. With these approximations, $\alpha$ is independent of $\log \tl$ and $\log \tc$ and can be ignored for fixed $\log \tobs$.

The probability distribution we have found for $\log t_{\rm
  c}$ and $\log t_{\Lambda}$ is a function of powers of $\tc$
and $t_{\Lambda}$, i.e., exponential in the logarithms.  Therefore the
distribution will be dominated by its maximum.  A useful way of
determining the location of the maximum is to follow the gradient flow
generated by the probability distribution.   In the language of
Ref.~\cite{HalNom07, BouHal09}, this is a multiverse force pushing
$\log \tl$ and $\log \tc$ to preferred values. We could use our
formulas to determine the precise direction of the multiverse force,
but this would be difficult to represent graphically, and it is not
necessary for the purpose of finding the maximum.  Instead, we shall
indicate only whether each of the two variables prefers to increase or
decrease (or neither), by displaying horizontal, vertical, or diagonal
arrows in the $(\log \tc, \log \tl)$ plane
(Fig.~\ref{fig-patchflow}). We ignore the prior $g(\log \tc)$ for now since it is not exponential.

We consider each region in \eqref{eq-pcppos}  in turn.  In region I, $\tobs < \tc < \tl$, the probability is proportional to $\tc^{-1}$. Hence, there is a pressure toward smaller $\tc$ and no pressure on $t_{\Lambda}$.  This is shown by a left-pointing horizontal arrow in region I of the figure.  In region II ($\tc<\tobs<t_{\Lambda}$), the probability is proportional to $\tc \tobs^{-2}$.  This pushes toward larger $\log \tc$ and is neutral with respect to $\log \tl$.  (Recall that we are holding $\log \tobs$ fixed for now.)  In region III ($\tc < t_{\Lambda} < \tobs)$, the probability goes like $\tc t_{\Lambda}^{-2} e^{-3 \tobs/t_{\Lambda}}$.  Since the exponential dominates, the force goes toward larger $\log t_{\Lambda}$; $\log \tc$ is pushed up as well.   In region IV ($\tl<\tobs,~\tl<\tc$) the exponential again dominates, giving a pressure toward large $\log t_{\Lambda}$. In region V ($\tobs<t_{\Lambda}<\tc$), the distribution is proportional to $\tl^{-1}$, giving a pressure toward small $\log t_{\Lambda}$. The dependence of the probability on $\log \tc$ lies entirely in the prior $g(\log \tc)$ in these last two regions because the universe is approximately flat and hence dynamically independent of $\log \tc$.

Leaving aside the effect of $g(\log \tc)$ for now, we recognize that the probability density in Fig.~\ref{fig-patchflow} is maximal along two lines of stability, $\log t_{\Lambda} = \log \tobs$ and $\log \tc = \log \tobs$, along which the probability force is zero.  These are the boundaries between regions IV/V and I/II, respectively.  They are shown by thick lines in Fig.~\ref{fig-patchflow}.  The fact that the distribution is flat along these lines indicates a mild runaway problem: at the crude level of approximation we have used thus far, the probability distribution is not integrable.  Let us consider each line in turn to see if this problem persists in a more careful treatment.

Along the line $\log t_{\Lambda} = \log \tobs$, the prior distribution $g(\log \tc)$ will suppress large $\log \tc$ and make the probability integrable, rendering a prediction for $\log \tc$ possible.   (With a plausible prior, the probability of observing a departure from flatness in a realistic future experiment is of order 10\%~\cite{FreKle05, DeSSal09}; see also~\cite{BouLei09}. See~\cite{MarRiv06, BozAlb09} for different priors.)  We will find the same line of stability for the other two measures in the following section, and it is lifted by the same argument.

The line $\log \tc = \log \tobs$ looks more serious.  The prior on $\log \tl$ follows from very general considerations~\cite{Wei87} and cannot be modified.  In fact, the probability distribution is rendered integrable if, as we will suppose, the landscape contains only a finite number ${\cal N}$ of vacua.  This implies that there is a ``discretuum limit'', a finite average gap between different possible values of $\Lambda$. This, in turn, implies that there is a smallest positive $\Lambda$ in the landscape, of order 
\begin{equation}
\Lambda_{\rm min}\sim 1/{\cal N}~.
\end{equation}
This argument, however, only renders the distribution integrable; it
does not suffice to bring it into agreement with observation.  It
tells us that $\log \Lambda$ is drawn entirely at random from values
between $\log \Lambda_{\rm min}$ and $\log \tobs^{-2}$, so at this level of analysis we do not predict the
coincidence $\log \tobs \approx \log \tl$.

Even though we do not predict a coincidence for observers living at a
fixed arbitrary $\log \tobs$, it could still be the case that after
averaging over the times when observers could live most observers see
$\log \tobs \sim \log \tl$.  To address this question, we need to allow $\log \tobs$
to vary.
For fixed $\log \tobs$, the maximum of the probability with respect to $\log \tc$
and $\log \tl$ is obtained along the boundary between
region I and region II, as discussed above. Along this line, the
probability is
\begin{equation} \label{eq-poscptobs}
\frac{dp}{d \log \tobs}
\sim {g(\log \tobs) \over \tobs} {\alpha(\log \tobs)} ~.
\end{equation}
Having maximized over $\log \tl$ and $\log \tc$, we can now ask at what $\log \tobs$
the probability is maximized. Note that since the distribution is
exponential, maximizing over $\log \tl$ and $\log \tc$ is the same as
integrating over them up to logarithmic corrections coming from the
function $g$.

The location of the maximum depends on the behavior of $\alpha(\log \tobs)$.  Let us assume that 
\begin{equation}
\alpha \sim \tobs^{1 + \eps}\ , \quad \mbox{with} \ \  \eps>0\, .
\label{eq-asdf}
\end{equation}
(We will justify this assumption at the end of this section, where we will also describe what happens if it is not satisfied.) Then the maximum is at the largest value of $\log \tobs$ subject to the constraint defining regions I and II, $\log \tobs <\log \tl$.  It follows that the maximum of the three-variable probability distribution is at
\begin{equation}
\log \tobs \approx \log \tc \approx \log \tl \approx \log \tl^{\rm max}~.
\label{eq-cpp}
\end{equation}
Therefore, in a landscape with $\eps>0$ and vacua with $\Lambda>0$, the causal patch predicts that all three scales are ultimately set by $\log \tl^{\rm max}$, and thus, by the (anthropic) vacuum with smallest cosmological constant.  This, in turn, is set by the discretuum limit, i.e., by the number of vacua in the landscape~\cite{Bou06,BouHar07,BouHal09,BouHar10}, according to
\begin{equation}
\tl^{\max} \sim {\bar{\cal N}}^{1/2} ~.
\end{equation}
This is a fascinating result~\cite{BouFre10c}.  It implies that the remarkable scales we observe in Nature, such as the the vacuum energy and the current age of the universe, are not only mutually correlated, but that their absolute scale can be explained in terms of the size of the landscape.  If current estimates of the number of vacua~\cite{BP,DenDou04b} hold up, i.e., if $\log_{10} {\cal
  N}$ is of order hundreds,\footnote{The number of anthropic vacua,
  $\bar{\cal N}$, may be smaller by dozens or even hundreds of orders
  of magnitude than the total number of vacua, $\cal N$, for
  low-energy reasons that are unrelated to the cosmological constant
  or curvature and so are not included in out analysis.  Hence,
  $\log_{10} {\cal N}\sim O(1000)$ may be compatible with
  Eq.~(\ref{eq-cpp}).} then Eq.~(\ref{eq-cpp}) may well prove to be in
agreement with the observed value $\tl \sim 10^{61}$.

Let us go somewhat beyond our order-of-magnitude estimates and determine how precisely $\log \tobs$ and $\log \tl$ can be expected to agree.  To that end, we will now calculate $p_{\rm CP}(f^{-1}\tl < \tobs < f \tl)$ as a function of $f$, i.e., the probability that $\log \tobs$ lies within an interval $\log f$ of $\log \tl$.
The probability distribution of Eq.~\ref{eq-pcppos} is dominated near the boundary of regions IV and V, and the probability in region IV is exponentially suppressed.  So we will neglect all regions except region V.  (Ignoring region IV means we are eliminating the possibility that $\log \tobs>\log \tl$.)

The probability density in region V is
\begin{equation}
\frac{dp}{d\log \tobs~d\log \tc~d\log \tl} \propto \frac{\tobs^{1+\eps}}{\tl}g(\log \tc)~.
\end{equation}
We will further restrict to $\tc > \tl^{\rm max}$, which is reasonable if $\tobs$ is pushed to large values and $g(\log \tc)$ does not strongly prefer small values of $\log \tc$.  Since we are computing a probability marginalized over $\log \tc$, this restriction on the range of $\log \tc$ means that the exact form of $g$ will not affect the answer.  The quantity
\begin{equation}
\int_{\log \tlmax}^\infty d\log \tc ~g(\log \tc)
\end{equation}
will factor out of our computations, and hence we will ignore it.
Having eliminated the $\log \tc$ dependence, we continue by computing the normalization factor $Z$ for $\log \tl >\log \tobs$:
\begin{equation}
Z = \int_0^{\log \tl^{\rm max}}d\log \tobs \int_{\log \tobs}^{\log \tl^{\rm max}} d\log \tl ~\frac{\tobs^{1+\eps}}{\tl}
\approx \frac{(\tl^{\rm max})^\eps}{\eps(1+\eps)}~.
\end{equation} 
In the last line we have dropped terms negligible for $\tl^{\rm max} \gg 1$.

Now we will calculate the unnormalized probabilty for $f^{-1}\tobs < \tl < f\tobs$.  We will split the integration region into two subregions according to whether $\tobs < f^{-1}\tl^{\rm max}$ or $f^{-1}\tl^{\rm max} < \tobs < \tl^{\rm max}$.  It turns out that each of these subregions is important.  First we do $\tobs < f^{-1}\tl^{\rm max}$:
\begin{align}
\int_0^{\log (f^{-1}\tl^{\rm max})}&d\log \tobs  \int_{\log \tobs}^{\log (f \tobs)} d\log \tl ~\frac{\tobs^{1+\eps}}{\tl}\\
&\approx \frac{(\tl^{\rm max})^\eps}{\eps}(f^{-\eps} - f^{-1-\eps})\\
&=Z (1+\eps)(f^{-\eps} - f^{1-\eps})~.
\end{align}
Finally we calculate the case $f^{-1}\tl^{\rm max} < \tobs < \tl^{\rm max}$:
\begin{align}
\int_{\log (f^{-1}\tl^{\rm max})}^{\log \tl^{\rm max}}&d\log \tobs \int_{\log \tobs}^{\log \tl^{\rm max}} d\log \tl ~\frac{\tobs^{1+\eps}}{\tl}\\
&= (\tlmax)^\eps \left[\frac{1-f^{-\eps}}{\eps}- \frac{1-f^{-1-\eps}}{1+\eps}\right]\\
&= Z\left[1 - (1+\eps)f^{-\eps}+\eps f^{-1-\eps})\right]~.
\end{align}

Adding together the unnormalized probabilities and dividing by the factor $Z$ we find the result
\begin{equation}
p_{\rm CP}(f^{-1}\tobs < \tl < f\tobs) \approx 1-f^{-1-\eps}~.
\end{equation}
In addition to being independent of $g(\log \tc)$, this result is independent of $\tl^{\rm max}$, but the validity of our approximation depends on both.  In particular, region V contributes more to the probability for larger $\tl^{\rm max}$, so the approximation gets better as $\tl^{\rm max}$ increases.  However, even for $\tl^{\rm max} = 10^{60}$ the result is only off by a few percent compared to a numerical integration.  

Let us now return to discussing our assumption, Eq.~(\ref{eq-asdf}).  If $\eps$ were not positive, that is, if $\alpha$ increased at most linearly with $\tobs$, then the maximum of the probability distribution would be located at the smallest value of $\tobs$ compatible with observers.  In this case the causal patch would predict $t_\Lambda \gg t_{\rm obs}$.  This would be in conflict with observation except under the extremely contrived assumption that $t_\Lambda^{\rm max} \sim t_{\rm obs}^{\rm min}$.

However, the assumption that $\eps>0$ is quite plausible~\cite{BouFre10c}.  Recall that we are only discussing the form of $\alpha$ in the regime where $\tobs$ is the shortest time scale, $\tobs \lesssim \tc, \tl$, so we do not have to worry that later observations may be disrupted by curvature or vacuum energy.  Recall, moreover, that $\alpha$ is defined by averaging over many vacua, so we must consider only how this average depends on $t_{\rm obs}$.  In particular, this means that we should not imagine that in moving from one value of $t_{\rm obs}$ to another, we need to hold fixed the vacuum, or even restrict to only one or two parameters of particle physics and cosmology.  Typical vacua with most observers at one value of $t_{\rm obs}$ are likely to differ in many details from vacua in which most observers arise at a different time.  

With this in mind, we note two general effects that suggest that $\alpha(\log t_{\rm obs})$ increases monotonically.  First, the spontaneous formation of highly complex structures such as observers relies both on chance and, presumably, on a long chain of evolutionary processes building up increasing complexity.  The later the time, the more likely it is that such a chain has been completed.  Secondly, for larger $\tobs$, the same amount of mass can be distributed among more quanta, of less energy each.  Therefore, less mass is necessary to construct a system containing a given number of quanta, such as a system of sufficient complexity to function as an observer.  These arguments make it very plausible that $\alpha$ grows.  Moreover, while they do not prove that it grows more strongly than linearly with $t_{\rm obs}$, they do make this type of behavior rather plausible.

\subsection{Negative cosmological constant}
\label{sec-cpneg}

We turn to negative values of the cosmological constant, beginning with the case $\Lambda<0$, $\tc \ll \tl$.  From Eqs.~(\ref{eq-negsf}) and (\ref{eq:eh}), we find that the comoving radius of the causal patch is given by
\begin{equation} \label{eq:eh_n}
\chi_{\rm CP}(t)\sim \left\{\begin{array}{ll}
3- 2 \log(\tc/2t_{\Lambda}) + 3\left[1-(t/\tc)^{1/3}\right]
~,& t<\tc \\
3 -\log\tan (t/2 \tl) +\log\tan (\tc'/2 \tl)~,& \tc<t<\tc' \\
3 \left(\frac{t'}{\tc}\right)^{1/3}~, & \tc'<t~.
\end{array}\right.
\end{equation}
Recall that a prime denotes time remaining before the crunch: $t'\equiv t_{\rm f}-t$.  To better show the structure of the above expressions, we have kept some order-one factors and subleading terms that will be dropped below.  We will approximate $\log \tan (\tc'/2t_{\Lambda}) = -\log\tan (\tc/2t_{\Lambda}) \approx - \log (\tc/2t_{\Lambda})$.

The mass inside the causal patch at the time $\tobs$ is
\begin{equation}
M_{\rm CP} = \rho a^3 V_{\rm CP}[\chi_{\rm CP}(\tobs)] \sim \tc V_{\rm CP}~.
\end{equation}
We will again approximate the comoving volume inside a sphere of
radius $\chi$ by $\chi^3$ for $\chi\lesssim 1$ and by $e^{2\chi}$ for
$\chi\gtrsim 1$, giving
\begin{equation} 
\mcp  \sim  \left\{\begin{array}{lll}
t_{\Lambda}^4 /\tc^{3}~, & \tobs<\tc  \ \ & I\\
\tl^2 \tc^{-1} \tan^{-2}(\tobs/2t_{\Lambda})~,& \tc<\tobs<\tc'\ \ & II \\
\tobs' , &  \tc' < \tobs \ \ &III
\end{array}\right.
\label{eq-mcpneg1}
\end{equation}

Now let us consider the case $\tc\gtrsim t_{\rm f}/2$.  The comoving radius of the causal patch is given by using Eqs.~(\ref{eq-approxnegflata}) and~(\ref{eq:eh}):
\begin{equation} 
\chi_{\rm CP}(t)\sim \left\{\begin{array}{ll}
(t^\prime)^{1/3} \tc^{-1/3}
~,& \tf/2<t  \\
2(\tf/2\tc)^{1/3} - (t/\tc)^{1/3}~, & t < \tf/2~.
\end{array}\right.
\end{equation}
The mass in the causal patch is then given by, up to order one constant factors,
\begin{equation} 
\mcp  \sim  \left\{\begin{array}{lll}
\tobs^\prime~, & \tf/2 < \tobs \ \ & IV\\
\tf , &  \tobs < \tf/2 \ \ &V
\end{array}\right.
\label{eq-mcpneg2}
\end{equation}

Now we can combine all of the above information to obtain the full
probability distribution, 
\begin{equation} 
\frac{d^3p_{\rm CP}}{d\log \tc~d\log \tl d \log \tobs} \sim  g \alpha \times\left\{\begin{array}{lll}
\vspace{.05in}\displaystyle{ \tl^2 \over \tc^3}~, & \tobs<\tc
< \tl \ \ & I\\
\vspace{.05in}
\displaystyle{1 \over \tc \tan^2(\tobs/2\tl)}~,&
\tc<\tobs< \tc^\prime~, \tc < \tl~,  \ \ & II \\
\vspace{.05in}\displaystyle{\tobs^\prime  \over \tl
  ^2}~, & \tc^\prime < \tobs~, \tc< \tl \ \ &III \\
\vspace{.05in} \displaystyle{\tobs^\prime \over \tl^2}~, & \tf/2 < \tobs < \tc \ \ & IV \\
\displaystyle{1 \over \tl}~, & \tobs < \tf/2 < \tc \ \ & V
\end{array}\right.
\label{eq-pcpneg}   
\end{equation}

The analysis of the probability ``forces'' proceeds as in the positive
cosmological constant case discussed in the previous subsection, by
identifying and following the directions along which the probability
grows in each distinct region of the $(\log \tl, \log \tc)$
plane. 
The result, however, is rather different
(Fig.~\ref{fig-patchflow}b). For fixed $\log \tobs$, the unnormalized probability density
diverges in the direction of small $\log \tc$ and large $\log t_{\Lambda}$
(region II)
like $\tl^2 \tc^{-1}$.  The discrete
spectrum of $\Lambda$ bounds $\log \tl$ from above, and the Planck
scale is a lower limit on $\log \tc$.  
Recall that so far, we have approximated the rate of observations per unit mass $\alpha$
as independent of $(\log \tc ,\log
\tl)$.   However, if
$\tc\ll \tobs$ ($\tl \ll \tobs$), then
curvature (or vacuum energy) could dynamically affect the processes by
which observers form.  One would expect that such effects are
generally detrimental. 

Here, for the first time, we find a distribution that peaks in a
regime where $\tc\ll \tobs$.  This means that the detailed
dependence of $\alpha$ on $\log \tc$ is important for understanding the
prediction and must be included.  We do not know this function except
for special classes of vacua.

Instead of letting $\log \tc \rightarrow 0$ so that $\tc$ becomes Planckian, we will only allow $\tc$ to fall as low as $\tc^{\rm min}$.  We do this because it does not make our analysis any more difficult, and it may capture some aspects of anthropic selection effects if we choose to set $\log \tc^{\rm min}$ to be some positive quantity.

Thus, within our current
approximations
 the causal patch predicts
that most observers in vacua with negative cosmological constant
measure
\begin{equation}
\log t_c\to \log \tc^{\rm min},~~ \log \tl \to \log \tl^{\rm max}~,
\end{equation}
where $\tl^{\rm max}\equiv |\Lambda|_{\rm min}^{-1/2}\sim {\cal
  N}^{1/2}$ is the largest achievable value of $\tl$ in the
landscape. 
Our result reveals a preference for separating the curvature, observer, and vacuum
timescales: a hierarchy, rather than a coincidence.

What happens if $\log \tobs$ is also allowed to vary?  After
optimizing $\log \tl $ and $\log \tc$, the probability distribution over
$\log \tobs $ is
\begin{equation}\label{eq-negcptobs}
{dp \over d \log \tobs} \sim \left( \tl^{\rm max} \over \tobs
\right)^2 \alpha(\log \tobs) \frac{g(\log \tc^{\rm min})}{\tc^{\rm min}}~.
\end{equation}
If $\alpha$ grows faster than quadratically in $\tobs$, then large values of $\log \tobs$ are
predicted: $\log \tobs\sim \log \tl \sim \log \tl^{\rm max}$,
$\log \tc\sim \log \tc^{\rm min}$, with the maximum probability density given by
$\alpha(\log \tl^{\rm max}) g(\log \tc^{\rm min})/\tc^{\rm min}$.  Otherwise, a small value of $\log \tobs$ is
predicted: $ \log \tl \sim \log \tl^{\rm max}$, $\log \tobs\sim \log \tobs^{\rm min}$,  $\log \tc\sim \log \tc^{\rm min}$, with maximum probability $(\tl^{\rm max}/\tobs^{\rm min})^2\alpha(\log \tobs^{\rm min}) g(\log \tc^{\rm min})/\tc^{\rm min}$~. (Here we have introduced $\log \tobs^{\rm min}$ in an analogous way to $\log \tc^{\rm min}$.  The point here is that typical observers live at the earliest possible time.)

Do these predictions conflict with observation?  Not so far: We are
observers in a vacuum with $\Lambda>0$, so the relevant probability
distribution over $(\log \tl, \log \tc )$ is the one computed in the
previous subsection.  This led to the predictions that $\log \tl\sim
\log \tobs$ and $\log \tc \gtrsim \log \tobs$, both of which agree
well with observation; and that the scale of $\log \tl$ is controlled
by the number of vacua in the landscape, which is not ruled out.

However, we do get a conflict with observation if we ask about the
total probability for each sign of the cosmological constant.  The
total probability for positive cosmological constant is approximately given by the
value of the distribution of the maximum. With our assumption about
$\alpha$ (Eq.~\ref{eq-asdf}), this is $p_+ \sim g(\log \tl^{\rm max}) \alpha(\log \tl^{\rm
  max})/\tl^{\rm max}$. The total probability
for negative $\Lambda$ is also controlled by the probability density
at the maximum of the distribution; as mentioned earlier, it is given
by $\alpha(\log \tl^{\rm max}) g(\log \tc^{\rm min})/\tc^{\rm min}$ if $p>1$, and by $(\tl^{\rm max}/\tobs^{\rm min})^2\alpha(\log \tobs^{\rm min}) g(\log \tc^{\rm min})/\tc^{\rm min}$
for $p<1$. 

Dividing these, we find that
a negative value of $\Lambda$ is favored by a factor
\begin{equation}
{p_- \over p_+} = \frac{\tl^{\rm max} g(\log \tc^{\rm min})}{\tc^{\rm min}g(\log \tlmax)}~ {\rm
  for}\; p > 1~.
\end{equation}
We know that $\tl^{\rm max}$ must be at least as large as the observed value
of $\tl$, which is of order $\tobs$: $\tl^{\rm
  max}>\tobs\sim 10^{61}$.  Furthermore, we expect that $g(\log
\tlmax) < g(\log \tc^{\rm min})$.  It follows that $p_+< \tc^{\rm min}
/\tl^{\rm max}$: the
observed sign of the cosmological constant is extremely unlikely
according to the causal patch measure in our simple model unless
$\tc^{\rm min}$ is rather close to $\tobs$. The
situation is similarly bad if $p < 1$.

We regard this result as further evidence~\cite{Sal09,BouLei09} that
the causal patch cannot be applied in regions with nonpositive
cosmological constant, or more generally, in the domains of dependence
of future spacelike singularities and hats.  This is plausible in
light of its relation to the light-cone time
cutoff~\cite{Bou09,BouFre10}, which is well-motivated~\cite{GarVil08}
by an analogy to the UV/IR relation~\cite{SusWit98} of the AdS/CFT
correspondence~\cite{Mal97}, but only in eternally inflating regions.

\section{The apparent horizon cutoff}
\label{sec-ah}

This section is structured like the previous one, but we now consider the apparent horizon cutoff, which is introduced here for the first time.  

\subsection{Definition}
\label{sec-ahdef}

\begin{figure}[tbp]
\centering
 \includegraphics[width=5.in]{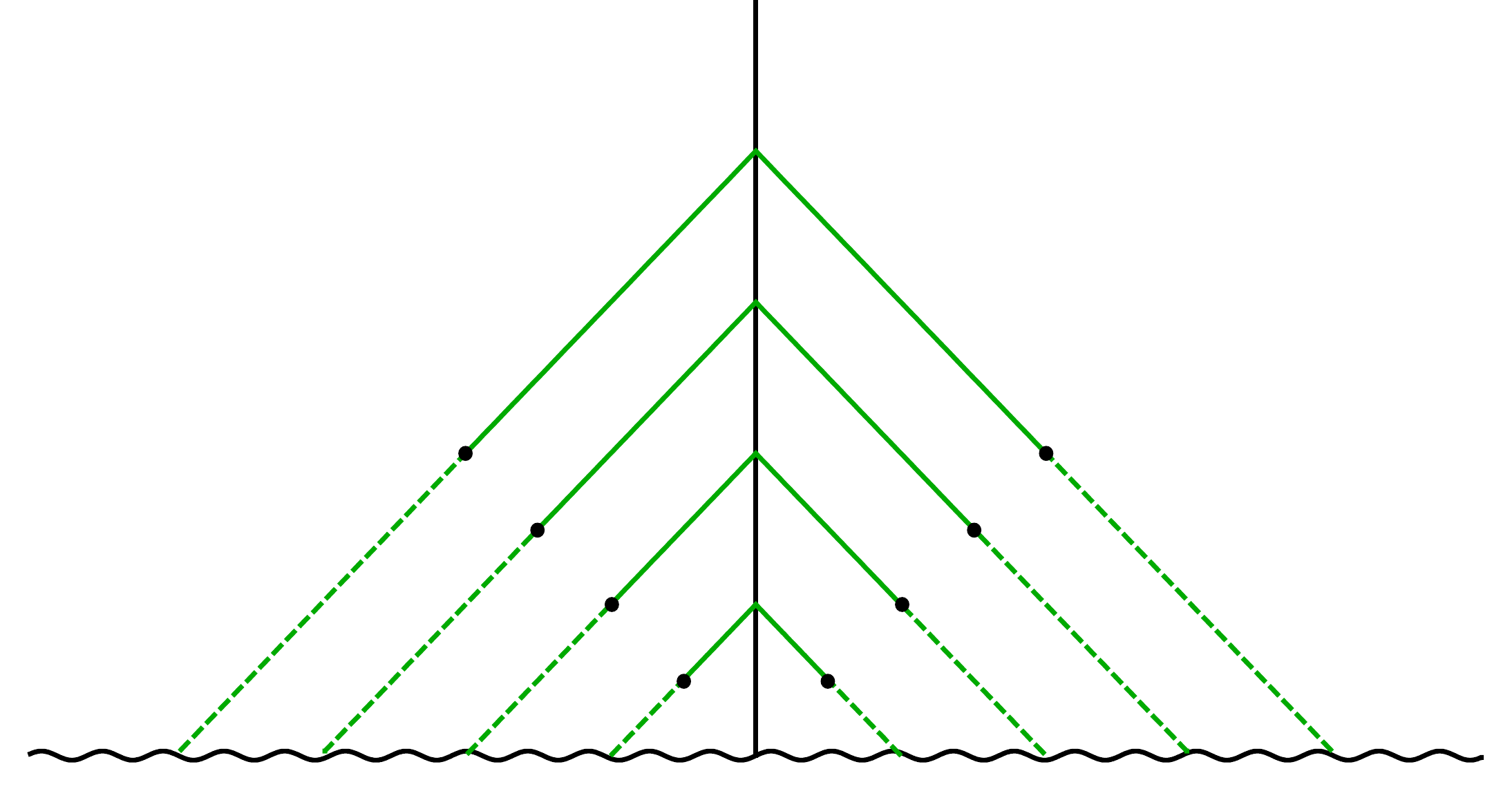}
   \caption{The causal patch can be characterized as the union of all past light-cones (all green lines, including dashed) of the events along a worldline (vertical line).  The apparent horizon cutoff makes a further restriction to the portion of each past light-cone which is expanding toward the past (solid green lines).  The dot on each light-cone marks the apparent horizon: the cross-section of maximum area, where expansion turns over to contraction.}
   \label{fig-appcut} 
\end{figure}

To define this cutoff, let us begin with a reformulation of the causal patch.  We defined the causal patch as the causal past of a point on the future boundary of spacetime.  But it can equivalently be characterized in terms of a worldline that ends on that point: the causal patch is the union of the past lightcones of all events that constitute the worldline. By the past light-cone we mean the null hypersurface that forms the boundary of the causal past.

Each past light-cone can be uniquely divided into two portions.  Beginning at the tip, the cross-sectional area initially expands towards the past.  But along each null geodesic generator of the light-cone, the expansion eventually becomes negative, and the cross-sectional area begins to decrease.  This turnaround does not happen in all spacetimes, but it does happen in any FRW universe that starts from a big bang (Fig.~\ref{fig-appcut}) or by bubble nucleation in a vacuum of higher energy. The point along each null geodesic where the expansion vanishes and the area is maximal is called the apparent horizon~\cite{RMP}.  The causal patch consists of both portions of the past light-cone.  The apparent horizon cutoff entails a further restriction: it consists only of the portion of each light-cone which is expanding towards the past.  

Our motivation for considering this cutoff comes from the preferred role played by the apparent horizon in understanding the holographic properties of cosmological spacetimes.  In the terminology of Refs.~\cite{CEB1,CEB2}, the apparent horizon is a {\em preferred holographic screen}: it possesses two light-sheets going in opposite spacetime directions, which together form an entire light-cone.  The covariant entropy bound states that any light-sheet off of a surface of area $A$ contains matter with entropy $S\leq A/4$.  Since the past light-cone consists of two different light-sheets off of the same surface of area $A_{\rm AH}$, the entropy on it cannot exceed $A_{\rm AH}/4+A_{\rm AH}/4=A_{\rm AH}/2$.  Both the causal patch cutoff and the apparent horizon cutoff can be thought of as a restriction to the information contained on the preferred holographic screen.  The causal patch keeps information about both sides of the screen; the apparent horizon cutoff only about one side.

The above definition of the apparent horizon cutoff applies to arbitrary worldlines in general spacetimes.  To obtain a definite ensemble of cutoff regions that can be averaged, let us specify that we follow geodesics orthogonal to an initial hypersurface specified according to some rule, for example, a region occupied by the longest lived de~Sitter vacuum in the landscape~\cite{GarSch05,BouYan07}.  When a geodesic enters a new bubble, it quickly becomes comoving~\cite{BouFre08b}.

\begin{figure}[tbp]
\centering
\subfigure[$\Lambda>0$]{
   \includegraphics[width=2.5in]{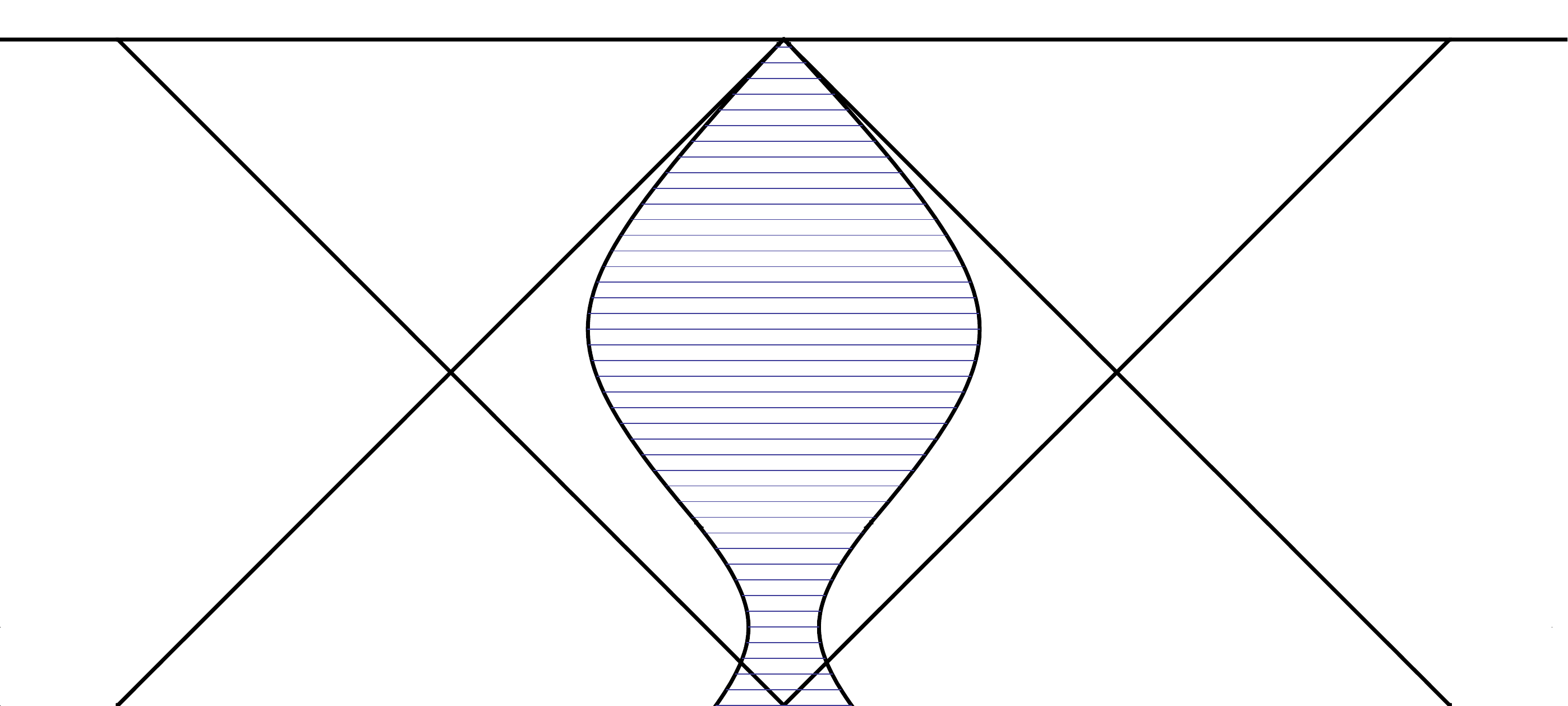}
   }
   \hspace{.5in}
    \subfigure[$\Lambda<0$]{
    \includegraphics[width=2.5 in]{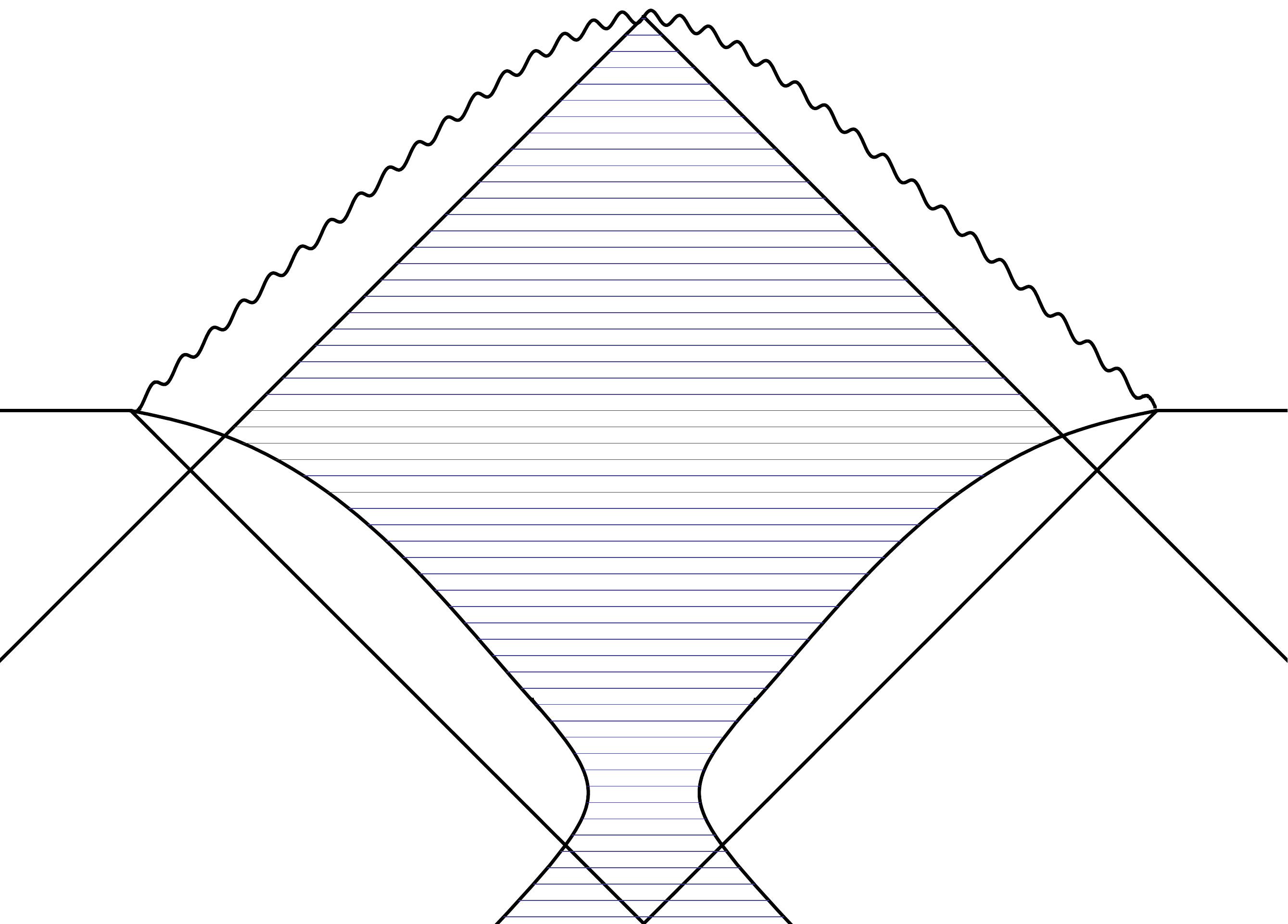}
    }
   \caption{
Conformal diagrams showing the apparent horizon cutoff region.  The boundary of the causal patch is shown as the past light-cone from a point on the conformal boundary. The domain wall surrounding a bubble universe is shown as the future light-cone of the bubble nucleation event.  The region selected by the cutoff is shaded.  For $\Lambda>0$ (a), the boundary of the causal patch is always exterior to the apparent horizon.  For $\Lambda<0$ (b), the apparent horizon diverges at a finite time.  Because the apparent horizon cutoff is constructed from light-cones, however, it remains finite.  The upper portion of its boundary coincides with that of the causal patch.}
   \label{fig-ah} 
\end{figure}

For a comoving geodesic in an FRW universe, it is convenient to restate the cutoff by specifying what portion of each FRW time slice should be included.   The apparent horizon at equal FRW time is defined as the sphere centered on the geodesic whose orthogonal future-directed ingoing light-rays have vanishing expansion.   This sphere exists at the FRW time $t$ if and only if the total energy density is positive, $\rho(t)>0$.  Its surface area is given by~\cite{CEB1}
\begin{equation}
A_{\rm AH}(t)=\frac{3}{2\rho(t)}~,
\label{eq-a1}
\end{equation}
from which its comoving radius can easily be deduced.  The apparent horizon cutoff consists of the set of points that are both within this sphere (if it exists), and within the causal patch.  The former restriction is always stronger than the latter in universes with positive cosmological constant, where the apparent horizon is necessarily contained within the causal patch~\cite{CEB2}.  In universes with $\Lambda<0$, there is an FRW time $t_*$ when the apparent horizon coincides with the boundary of the causal patch.  If $\tobs<t_*$, we restrict our attention to observers within the apparent horizon; otherwise we restrict to observers within the causal patch (see Fig.~\ref{fig-ah}).

\subsection{Positive cosmological constant}
\label{sec-ahpos}

We begin with the case $\Lambda>0$, $\tc \ll \tl$.  The scale factor $a(t)$ is given by Eq.~(\ref{eq:a}).   The energy density of the vacuum, $\rho_\Lambda\sim \Lambda\sim \tl^{-2}$, begins to dominate over the density of matter, $\rho_m\sim \tc/a^3$, at the intermediate time
\begin{equation}
t_i\sim \tc^{1/3} \tl^{2/3}~.
\end{equation}
Note that $\tc\ll t_i\ll \tl $ if $\tc$ and $\tl$ are well-separated.  Thus we can approximate Eq.~(\ref{eq-a1}) by
\begin{equation}
A_{\rm AH}(t)= \frac{3}{2(\rho_m(t)+\rho_\Lambda)}\sim \left\{\begin{array}{ll}
\rho_m^{-1}(t) ~, & t< t_i \\
\rho_\Lambda^{-1}~, & t> t_i
\end{array}\right.~.
\label{eq-a2}
\end{equation}
The comoving area of the apparent horizon, $A_{\rm AH}/a^2$, is initially small and grows to about one at the time $\tc$.  It remains larger than unity until the time $\tl$ and then becomes small again.
The  proper volume within the apparent horizon is $V_{\rm AH}\sim aA_{\rm AH}$ when the comoving area is large and $V_{\rm AH}\sim A_{\rm AH}^{3/2}$ when it is small.  
The mass within the apparent horizon is $M_{\rm AH}= \rho_mV_{\rm AH} \sim \tc V_{\rm AH}/a^3$.  Combining the above results, we find
\begin{equation} 
\label{eq:probah}
\mah \sim 
\left\{\begin{array}{lll}
\tobs~, & \tobs< t_i <\tl \ \ & I \\
\tc \tl^2 /\tobs^2~, & t_i<\tobs<\tl  \ \ & II \\
\tc e^{-3(\tobs/\tl -1)}~, & t_i< \tl < \tobs \  \ & III
\end{array}\right.~
\end{equation}

For the case $\Lambda>0$, $\tl\lesssim \tc$, the mass can be obtained by setting $\tc \sim  t_i \sim \tl$ in the above result:
\begin{equation} \label{eq-flatposahprob}
\mah \sim \left\{\begin{array}{lll}
\tobs~, & \tobs< \tl < \tc \ \ & V \\
\tl e^{-3(\tobs/\tl -1)}~, & \tl < \tobs,~ \tc\  \ & IV
\end{array}\right.~
\end{equation}

The full probability distribution is obtained as before by multiplying
by $\tobs \alpha(\log \tobs)$ and dividing by $\tl^2$ to get

\begin{equation} 
\frac{d^3p_{\rm AH}}{d\log \tc~d\log \tl d \log \tobs}
\sim g \alpha \times \left\{\begin{array}{lll}
\vspace{.05in}\displaystyle{\tobs \over \tl^2 }~, & \tobs<t_i
< \tl \ \ & I\\
\vspace{.05in}
\displaystyle{\tc \over \tobs^2}~,& t_i<\tobs<\tl  \ \ & II \\
\vspace{.05in}\displaystyle{\tc  \over
  \tl^2} 
\exp \left[-3 \left( \displaystyle{ 
      \tobs \over \tl} - 1\right)\right] ~, & t_i < \tl< \tobs \ \ &III \\
\vspace{.05in}\displaystyle{1 \over \tl} \exp
\left[-3 \left( \displaystyle{ 
      \tobs \over \tl} - 1 \right)\right]~, & \tl < \tobs , \tc \ \
& IV \\
\displaystyle{\tobs  \over \tl^2}~, & \tobs < \tl < \tc \ \ & V 
\end{array}\right.
\label{eq-pahpos}   
\end{equation}

\begin{figure}[tbp]
\centering
\subfigure[$\Lambda>0$]{
   \includegraphics[width=2.5in]{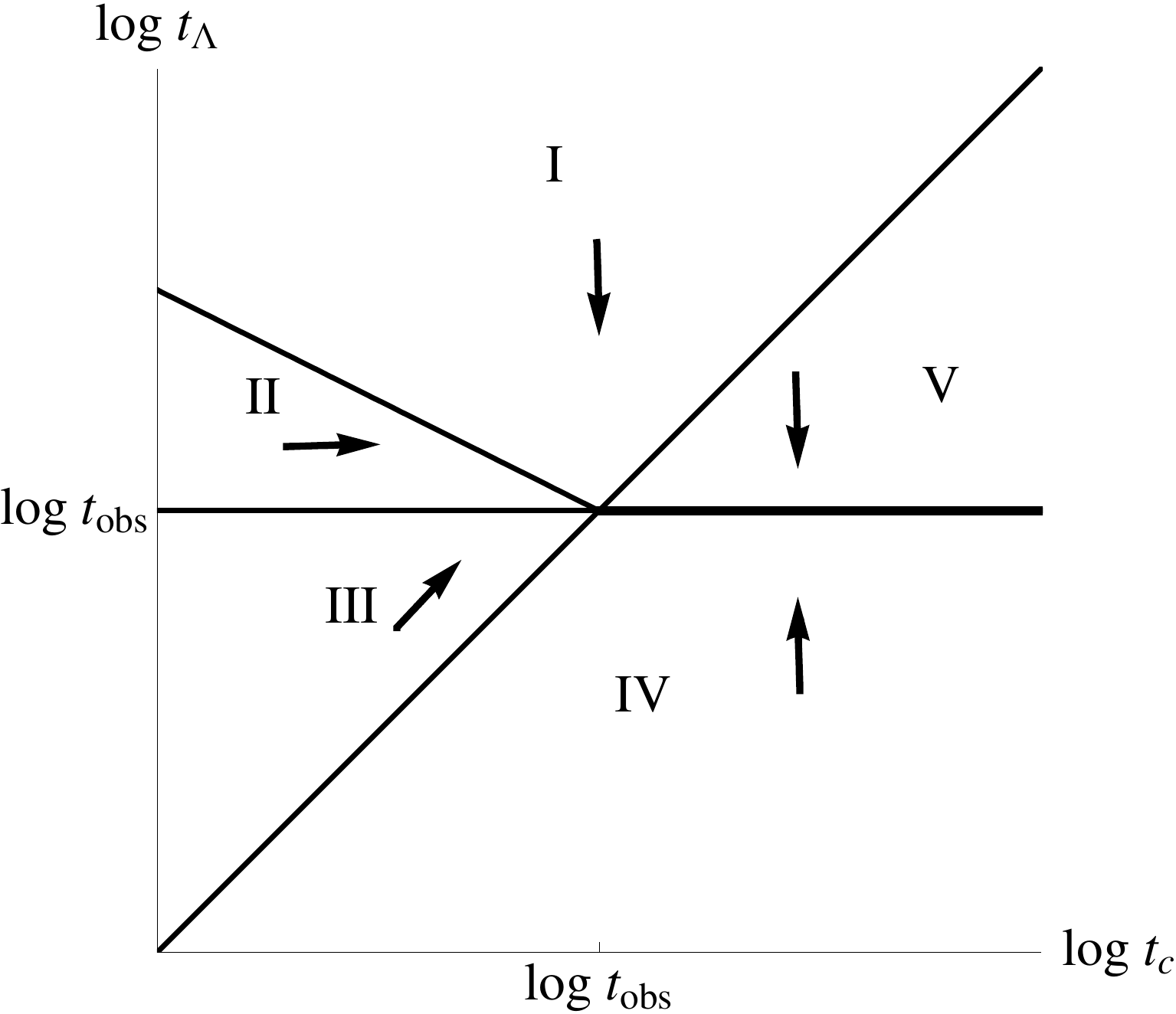}
   }
   \hspace{.5in}
    \subfigure[$\Lambda<0$]{
    \includegraphics[width=2.5 in]{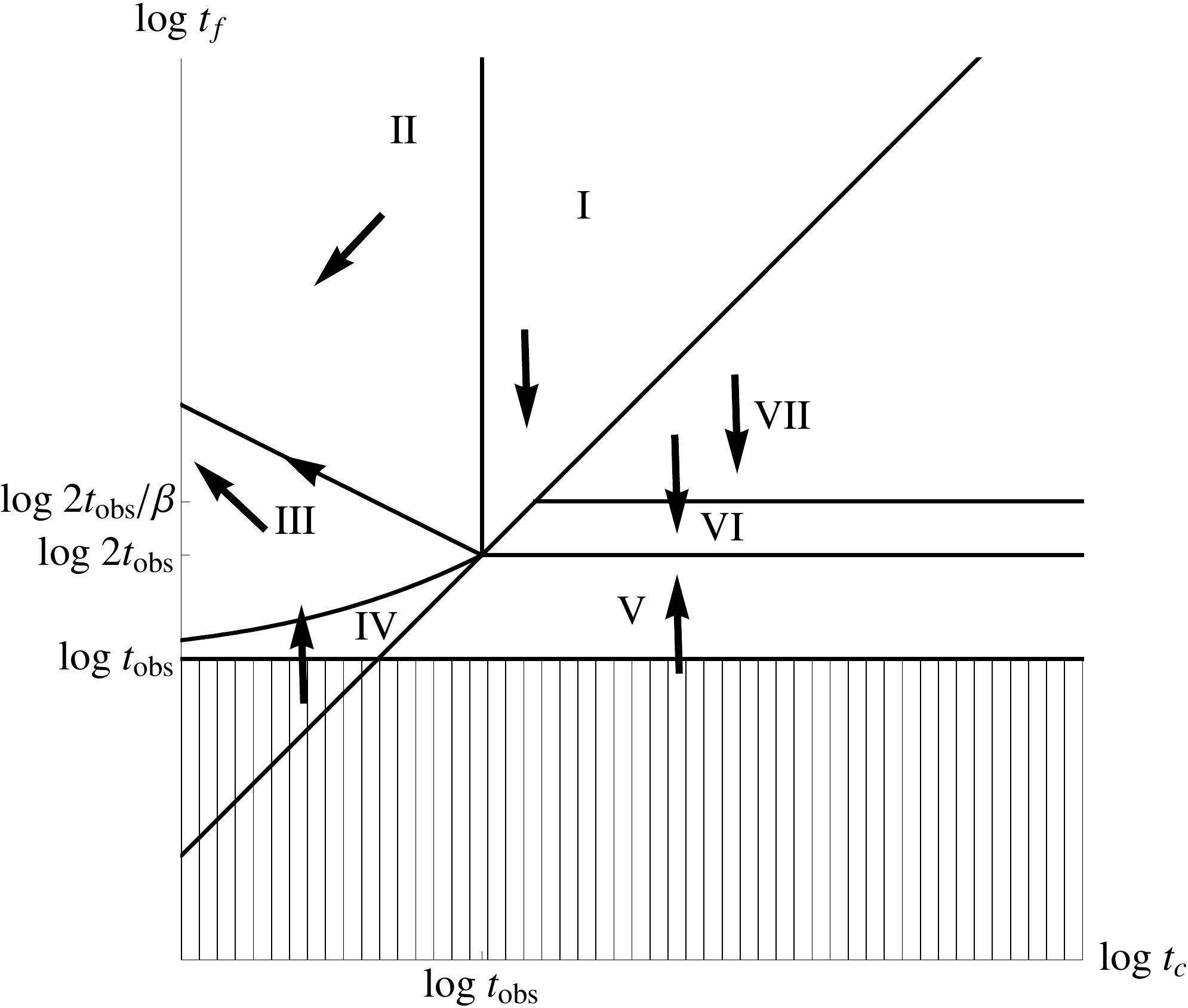}
    }
   \caption{The probability distribution from the apparent horizon cutoff.  The arrows indicate directions of increasing probability.  For $\Lambda>0$ (a), the probability is maximal along the boundary between regions IV and V before a prior distribution over $\log \tc $ is included.  Assuming that large values of $\tc$ are disfavored, this leads to the prediction $\log \tl \sim \log \tc \sim \log \tobs$.  For $\Lambda<0$ (b), the distribution is dominated by a runaway toward small $\tc$ and large $\tl$ along the boundary between regions II and III.}
   \label{fig-AHflow} 
\end{figure}
The probability forces are shown in Fig. \ref{fig-AHflow}. The boundary between regions I and II is given by $\log t_i = \log \tobs$, which corresponds to $\log t_{\Lambda} = \frac{3}{2}\log\tobs- \frac{1}{2} \log \tc$.  In region I, the probability is proportional to $t_{\Lambda}^{-2}$, corresponding to a force toward smaller $\log t_{\Lambda}$.  In region II there is a force toward large $\log \tc$.  In region III, the exponential dominates the $\log \tl$ dependence, giving a preference for large $\log t_{\Lambda}$; the $\tc$ prefactor provides a force towards large $\log \tc$. In regions IV and V the probabilities are independent of $\log \tc$ except for the prior $g(\log \tc)$.  The force is towards large $\log \tl$ in region IV, while in region V small $\log t_{\Lambda}$ is preferred.

Following the gradients in each region, we find that the distribution peaks on the boundary between regions IV and V.  Along this line, the probability density is constant except for $g(\log \tc)$.  As discussed in Sec.~\ref{sec-cppos}, this degeneracy is lifted by a realistic prior that mildly disfavors large values of $\log \tc$.  Thus, the apparent horizon cutoff predicts the double coincidence
\begin{equation}
\log \tobs\sim \log \tl\sim \log \tc~. 
\end{equation}
This is in good agreement with observation.

What if the observer time scale is allowed to vary? After optimizing
$\log \tl$ and $\log \tc$, the probability distribution over $\log \tobs$ is
\begin{equation}
\frac{dp}{d\log \tobs} \sim g(\log \tobs){\alpha(\log \tobs) \over \tobs}~.
\label{eq-jkl}
\end{equation}
We have argued in Sec.~\ref{sec-cppos} that $\alpha$ grows faster than $\tobs$; under this assumption, all three timescales are driven to the discretuum limit:
\begin{equation}
\log \tobs \approx \log \tc \approx \log \tl \approx \frac{1}{2}\log {\cal \bar{N}}~.
\end{equation}

\subsection{Negative cosmological constant}
\label{sec-ahneg}

We turn to the case $\Lambda<0$, $\tc\ll \tl$.  The scale factor is given by (\ref{eq-negsf}). The total energy density becomes negative at the intermediate time
\begin{equation}
t_i\sim \tc^{1/3} \tl^{2/3}~,
\label{eq-int}
\end{equation}
when the positive matter density is sufficiently dilute to be overwhelmed by the negative vacuum energy, $\rho_\Lambda\sim -\tl^{-2}$.  As discussed in Sec.~\ref{sec-ahdef}, the apparent horizon exists on the FRW timeslice $t$ only if the total density at that time is positive.  By Eq.~(\ref{eq-a1}), the apparent horizon diverges when the density vanishes.  Slightly earlier, at the time $t_* = (1-\epsilon)t_i$, the apparent horizon intersects the boundary of the causal patch.  For $t<t_*$, the apparent horizon and defines the cutoff; for $t>t_*$, the causal patch does (see Fig.~\ref{fig-ah}).

To compute $t_*$, notice that $\tc\ll \tl$ and Eq.~(\ref{eq-int}) imply $\tc\ll t_i\ll \tl$.  This implies that the scale factor can be well approximated by $a(t) \approx t_{\Lambda} \sin (t/t_{\Lambda}) \approx t$ in a neighborhood of $t_i$.  This range includes the time $t_*$ if $\epsilon$ is small.  We will assume this approximation for now, and we will find that $\epsilon\ll 1$ follows self-consistently.  By Eq.~(\ref{eq-a1}), the proper area of the apparent horizon at the time $t_i(1-\epsilon)$ is  $A_{\rm AH}(t) = \tl^2/2\epsilon$. From Eq.~\ref{eq:eh_n}, we find that the causal patch has proper area $16\pi e^3 \tl^4/\tc^2+O(\epsilon^2)$.  Equating these expressions, we find 
\begin{equation}
\epsilon = \frac{1}{32\pi e^3}\,\frac{\tc^2}{\tl^2}~,
\label{eq-epsilon}
\end{equation}
which is much less than unity.

For times $t< t_*$, we compute the mass within the apparent horizon.  When $t\lesssim 
\tc$ we use that $V_{\rm AH} \sim A_{\rm AH}^{3/2}$, while for $\tc\lesssim t < t_*$, we have $V_{\rm AH} \sim a A_{\rm AH}$. For times $t>t_*$ we use the results for the causal patch from Sec.~\ref{sec-cpneg}.  
\begin{equation}
\mah \sim \left\{\begin{array}{lll} 
\tobs, & \tobs< \tc \ \ & I\\
\tobs\left(1- (\frac{\tobs}{t_i})^3\right)^{-1}, & \tc<\tobs<t_* \ \ & II \\
\tl^2 \tc^{-1} \tan^{-2}(\tobs/2 t_{\Lambda}), & t_* < \tobs < \tc' \ \ & III \\
\tobs'~, & \tc'<\tobs \ \ & IV
\end{array}\right.~,
\label{eq-negprob}
\end{equation}

Finally, we consider the case $\Lambda<0$, $t_{\rm f}/2 < \tc$, for which the universe can be approximated as spatially flat at all times.  The scale factor is given by Eq.~(\ref{eq-negflata}).  The area of the apparent horizon, $A_{\rm AH} \sim \tl^2 \tan^2(\pi t/t_{\rm f})$, diverges at the turnaround time.  So at a time $t_*<t_{\rm f}/2$, the apparent horizon and causal patch are equal in size, and after that time we must use the causal patch as our cutoff.  The area of the causal patch is $A_{\rm CP} \sim \tl^2$ around this time, so the apparent horizon interesects the causal patch at
\begin{equation}
t_*^{\rm flat} \approx \beta {\tf \over 2}
\end{equation}
for $\beta$ some order one number less than one.

The comoving size of the apparent horizon is given by $\chi \sim (t/\tc)^{1/3}$ for $t < t_*^{\rm flat}$; for $t> t_*^{\rm flat}$ we use our formulas from the causal patch in the previous section to obtain
\begin{equation}
\mah \sim \left\{\begin{array}{lll} 
\tobs^\prime, & \tf/2 < \tobs < \tc \ \ & V\\
\tf~, & t_*^{\rm flat}< \tobs<\tf/2<\tc \ \ & VI \\
\tobs~, & \tobs < t_*^{\rm flat} < \tc \ \ & VII
\end{array}\right.~,
\end{equation}
We can now write the full probability distribution for the apparent horizon cutoff with negative cosmological constant,
\begin{equation} 
\frac{d^3p_{\rm AH}}{d\log \tc~d\log \tl d \log \tobs}
\sim g \alpha \times \left\{\begin{array}{lll}
\vspace{.05in}\displaystyle{\tobs  \over \tl^2 }~, & \tobs<t_c
< \tl \ \ & I\\
\vspace{.05in}
\displaystyle{ \tobs \over \tl^2 \left[ 1 - \left(\tobs \over t_i\right)^3\right] }~,& t_c<\tobs<t_* < \tl  \ \ & II \\
\vspace{.05in}\displaystyle{1 \over \tc \tan^2({\tobs \over 2 \tl})} 
~, & t_* < \tobs < \tc^\prime \ \ &III \\
\vspace{.05in}\displaystyle{\tobs^\prime  \over \tl^2}~, & \tc^\prime < \tobs < t_{\rm f} \ \
& IV \\
\vspace{.05in} \displaystyle{\tobs^\prime  \over \tl^2} & \tf/2 < \tobs< \tc &V \\
\vspace{.05in}\displaystyle{ 1 \over \tl}~, & t_*^{\rm flat}<\tobs < \tf/2 < \tc \ \ & VI \\
\displaystyle{\tobs \over \tl^2}~, & \tobs < t_*^{\rm flat} < \tc \ \ & VII  
\end{array}\right.
\label{eq-pahneg}   
\end{equation}

The probability force diagram is shown in Fig. \ref{fig-AHflow}. Just
looking at the arrows, it is clear that the maximum of the probability
distribution lies somewhere in region III, perhaps at the boundary
with region II. Although the formula in region III is already
reasonably simple, there is a simpler form that is correct at the same
level of approximation as the rest of our analysis,
\begin{equation}
 {\tl^2  \over \tc \tobs^2 }\alpha(\log \tobs) g(\log \tobs)~.
\end{equation}
This is a good approximation for $\tobs \ll \tl$, but it is only wrong
by an order one factor througout region III, so we will go ahead and
use this.

For fixed $\log \tobs$, it is clear that $\log \tl$ wants to be as large as
possible, and $\log \tc$ as small as possible, but we must remain in region
III. The condition $t_* < \tobs$ bounding region III is equivalent to
\begin{equation}
\log \tc + 2 \log \tl < 3 \log \tobs ~.
\end{equation}
If $\log \tobs$ is big enough so that $ \tc^{\rm min} (\tl^{\rm max})^2 <
\tobs^3$, then the maximum of the distribution is at $\log \tl = \log \tl^{\rm
  max}$ and $\log \tc = \log \tc^{\rm min}$,  with probability given by
\begin{equation}
{(\tl^{\rm max})^2 \over \tc^{\rm min} \tobs^2} \alpha(\log \tobs) g(\log \tobs) ~.
\label{bigtobs}
\end{equation}
If $\log \tobs$ is smaller, then the maximum is given by $\log \tc =\log  \tc^{\rm
  min}$ and $2\log \tl = 3\log \tobs - \log \tc^{\rm min}$, with probability
\begin{equation}
{\tobs \over (\tc^{\rm min})^2} \alpha(\log \tobs) g(\log \tobs) ~.
\label{smalltobs}
\end{equation}
In either case, we are driven to $\tc \ll \tobs$.

Note that, as in the case of the causal patch cutoff with $\Lambda<0$, the distribution is peaked in a regime where $\tc \ll \tobs$.  So there is some uncertainty in our result coming from the dependence of $\alpha$ on $\log \tc$ when $\log \tc < \log \tobs$.  We do not know the form of this function, which depends on details of the nature of observers, and as before we will just continue to assume that $\alpha$ is independent of $\log \tc$.



Now we allow $\log \tobs$ to vary. For small $\log \tobs$ such that
\eqref{smalltobs} is valid, $\log \tobs$ wants to grow given very mild
assumptions about $\alpha$. Eventually $\log \tobs$ becomes large enough
that we leave the small $\log \tobs$ regime. For larger $\log \tobs$ such that
\eqref{bigtobs} is valid, $\log \tobs$ is driven up to $\log \tl^{\rm max}$ if
$\alpha$ increases faster than quadratically with $\tobs$. In this case we
predict $\log \tl \sim\log\tobs$. If $\alpha$ grows more slowly with $\log \tobs$,
then we predict $\log \tobs \ll \log \tl$.

Let us compare the total probability for negative $\Lambda$ to the total probability for positive $\Lambda$, assuming the form \eqref{eq-asdf} for $\alpha$.  For negative $\Lambda$, we will assume that the large $\log \tobs$ regime is the relevant one, so that the correct probability distribution over $t_{\rm obs}$ is \eqref{bigtobs}.  Note that this is the the same as \eqref{eq-negcptobs}, the result for negative $\Lambda$ in the causal patch.
Additionally, \eqref{eq-jkl} is identical to \eqref{eq-poscptobs}, the result for positive $\Lambda$ in the causal patch.  So the total probabilities are identical to those we found previously for the causal patch.  Then
a negative value of $\Lambda$ is favored by a factor
\begin{equation}
{p_- \over p_+} = \frac{\tl^{\rm max} g(\log \tc^{\rm min})}{\tc^{\rm min}g(\log \tlmax)}~ {\rm
  for}\; p > 1~,
\end{equation}
and a similar result for $p<1$.

\section{The fat geodesic cutoff}
\label{sec-sf}

In this section, we compute probabilities using the fat geodesic cutoff, which considers a fixed proper volume $\Delta V$ near a timelike geodesic~\cite{BouFre08b}.  To compute probabilities, one averages over an ensemble of geodesics orthogonal to an initial hypersurface whose details will not matter.  As discussed in the previous section, geodesics quickly become comoving upon entering a bubble of new vacuum.  By the symmetries of open FRW universes, we may pick a fat geodesic at $\chi =0$, without loss of generality.

In the causal patch and apparent horizon measure, the cutoff region is large compared to the scale of inhomogeneities, which average out. The definition of the fat geodesic, however, is rigorous only if $\Delta V$ is taken to be infinitesimal.  Thus, in this section, we shall neglect the effects of local gravitational collapse.  We shall approximate the universe as expanding (and, for $\Lambda<0$ after the turnaround, contracting) homogeneously.  Since the physical 3-volume, $\Delta V$, of a fat geodesic is constant, the mass within the cutoff region is proportional to the matter density:
\begin{equation} 
M_{FG} \propto \rho_m\sim\frac{\tc }{a^3} ~.
\label{eq-mfg}
\end{equation}

The fat geodesic cutoff is closely related to the scale factor time cutoff, but it is more simply defined and easier to work with.  Scale factor time is defined using a congruence of timelike geodesics orthogonal to some initial hypersurface in the multiverse: $dt\equiv H d\tau$, where $\tau$ is the proper time along each geodesic and $3H$ is the local expansion of the congruence.  This definition breaks down in nonexpanding regions such as dark matter halos; attempts to overcome this limitation (e.g., Ref.~\cite{DGSV08}) remain somewhat ad-hoc.  In regions where the congruence is everywhere expanding, scale factor time is exactly equivalent to the fat geodesic cutoff with initial conditions in the longest lived de~Sitter vacuum~\cite{BouFre08b}.

\subsection{Positive cosmological constant}

We begin with the case $\Lambda > 0$, $\tc\ll \tl$.  Combining Eqs.~(\ref{eq-mfg}) and (\ref{eq:a}), we obtain
\begin{equation}
M_{\rm FG} \sim \left\{\begin{array}{lll} 
1/\tobs ^2, & \tobs< \tc < \tl \ \ & I\\
\tc/ \tobs^3 , & \tc<\tobs<t_{\Lambda}\ \ & II \\
(\tc/t_{\Lambda}^3) e^{-3\tobs/t_{\Lambda}}, & \tc<t_{\Lambda}< \tobs  \ \ & III 
\end{array}\right.~.
\end{equation}
For the flat universe ($\Lambda > 0$, $\tc> \tl$), we obtain
\begin{equation}
M_{\rm FG} \sim \left\{\begin{array}{lll} 
1/\tobs ^2, & \tobs< \tl \ \ & V\\
(1/t_{\Lambda}^2) e^{-3\tobs/t_{\Lambda}}, & t_{\Lambda}< \tobs\ \ & IV
\end{array}\right.~.
\end{equation}
This leads to the probability distribution
\begin{equation} 
\frac{d^3p_{\rm FG}}{d\log \tc~d\log \tl d \log \tobs}
\sim g \alpha \times \left\{\begin{array}{lll}
\vspace{.05in}\displaystyle{1 \over \tl^2 \tobs^2 }~, & \tobs<\tc
< \tl \ \ & I\\
\vspace{.05in}
\displaystyle{\tc \over \tobs^3 \tl^2}~,& \tc<\tobs<\tl  \ \ & II \\
\vspace{.05in}\displaystyle{\tc  \over
  \tl^5} 
\exp \left[-3 \displaystyle{ 
      \tobs \over \tl} \right] ~, & \tc < \tl< \tobs \ \ &III \\
\vspace{.05in}\displaystyle{1 \over \tl^4} \exp
\left[-3 \displaystyle{ 
      \tobs \over \tl} \right]~, & \tl < \tobs , \tc \ \
& IV \\
\displaystyle{1  \over \tobs^2 \tl^2}~, & \tobs < \tl < \tc \ \ & V 
\end{array}\right.
\label{eq-pfgpos}   
\end{equation}

\begin{figure}[tbp]
\centering
\subfigure[$\Lambda>0$]{
   \includegraphics[width=2.5in]{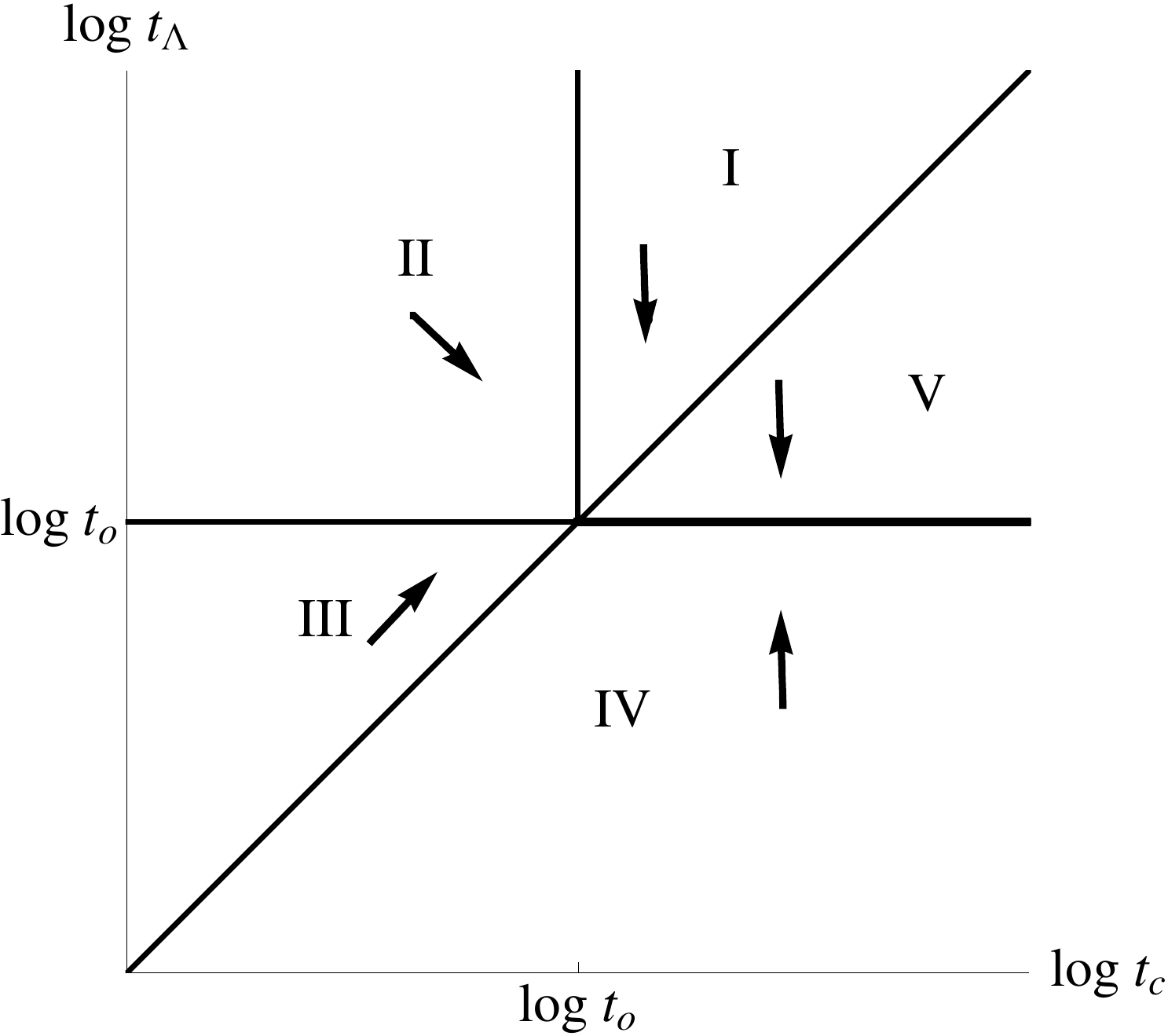}
   }
   \hspace{.5in}
    \subfigure[$\Lambda<0$]{
    \includegraphics[width=2.5 in]{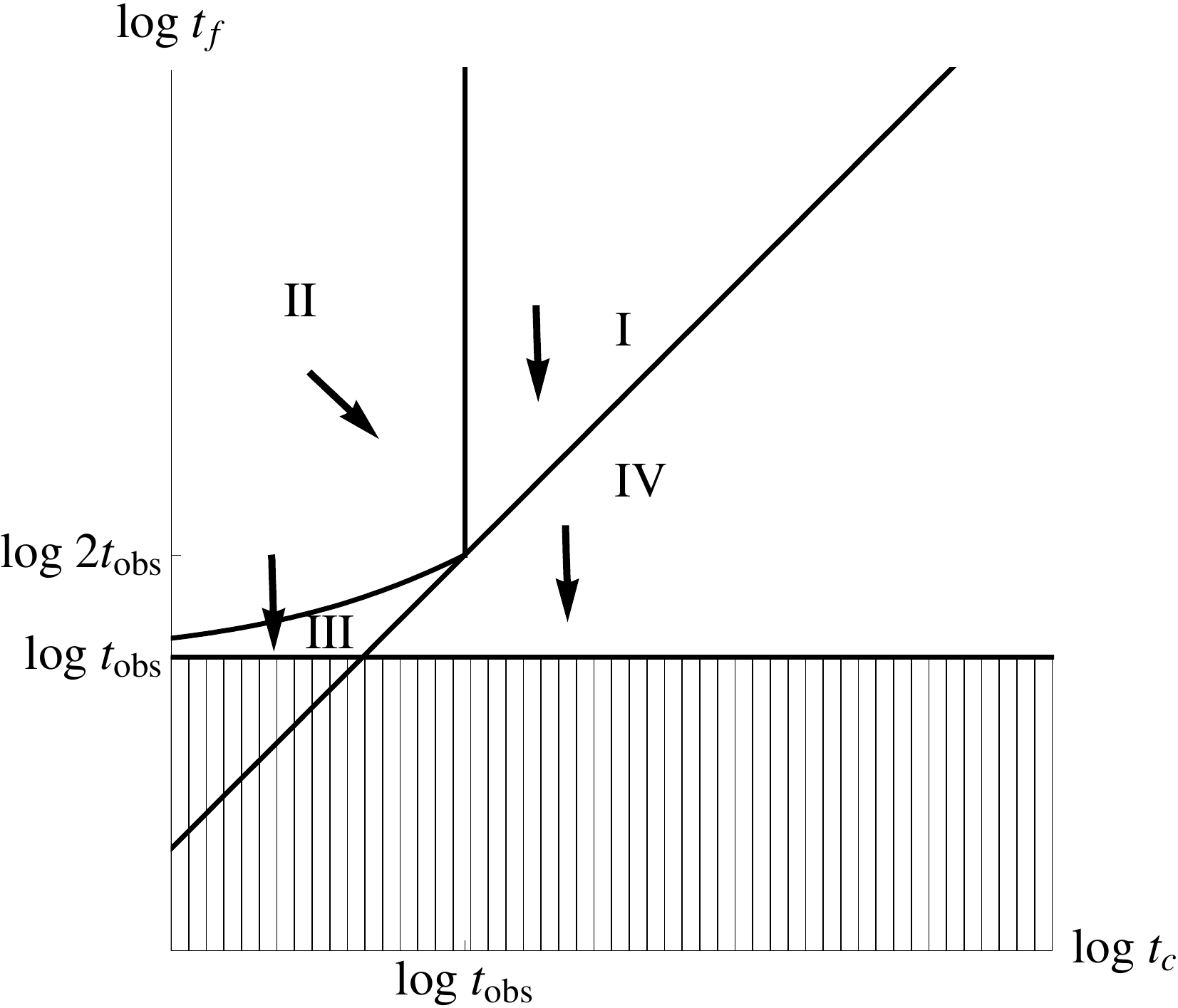}
    }
   \caption{The probability distribution computed from the scale factor (fat geodesic) cutoff.  The arrows indicate directions of increasing probability.  For $\Lambda>0$ (a), the probability distribution is maximal along the boundary between regions IV and V; with a mild prior favoring smaller $\log \tc$, this leads to the prediction of a nearly flat universe with $\log \tc \sim \log \tl \sim\log \tobs$.  For $\Lambda<0$ (b), the probability distribution diverges as the cosmological constant increases to a value that allows the observer timescale to coincide with the big crunch.}
   \label{fig-SFflow} 
\end{figure}

The probability force diagram is shown in Fig. \ref{fig-SFflow}.  The result is the same as for the apparent horizon cutoff: the distribution peaks on the entire line separating regions IV and V, up to the effects of $g(\log \tc)$.  A realistic prior that mildly disfavors large values of $\log \tc$ will tend to make $\log \tc$ smaller.  Thus, the fat geodesic cutoff predicts the double coincidence
\begin{equation}
\log \tobs\sim \log \tl\sim \log \tc~,
\end{equation}
in good agreement with observation.

What if we allow $\log \tobs$ to scan?  Optimizing $(\log
\tl,\log \tc)$, we find the probability distribution over $\log \tobs$:
\begin{equation}
\frac{dp}{d\log \tobs} \sim { \alpha(\log \tobs) g(\log \tobs) \over \tobs^4}~.
\label{eq-obssfp}
\end{equation}
The denominator provides a strong preference for $\log \tobs$ to be
small. To agree with observation, $\alpha$ must grow at least
like the fourth
 power of $\tobs$ for values of $\tobs$ smaller than the observed
 value $\tobs \sim 10^{61}$.
We cannot rule this out, but it is a much stronger assumption than the
ones needed for the causal patch and apparent horizon cutoffs.

The preference for early $\log \tobs$ in the fat geodesic cutoff can
be traced directly to the fact that the probability is proportional to
the matter density.  This result has an interesting
manifestation~\cite{BouFre08b} in the more restricted setting of
universes similar to our own: it is the origin of the strong
preference for large initial density contrast, $\delta \rho/\rho$,
which allows structure to form earlier and thus at higher average
density.

\subsection{Negative cosmological constant}
For $\Lambda<0$, $\tc\ll \tl$,  we use Eq.~(\ref{eq-negsf}) for the
scale factor.  The mass in the cutoff region is
\begin{equation}
M_{\rm FG}\sim \left\{\begin{array}{lll} 
1/\tobs^2, & \tobs< \tc \ \ & I\\
(\tc/t_{\Lambda}^3)\sin^{-3} (\tobs/t_{\Lambda}), & \tc<\tobs<\tc'\ \ & II \\
1/\tobs'^2, & \tc'<\tobs  \ \ & III 
\end{array}\right.~.
\label{eq-negsfprobcurv}
\end{equation}
(Recall that a prime denotes the time remaining before the crunch, $t'\equiv t_{\rm f}-t$.)  For the flat universe case, $\Lambda<0$ and $\tc>t_{\rm f}/2$, we use Eq.~(\ref{eq-negflata}) for the scale factor and find 
\begin{equation}\label{eq-negsfprob}
M_{\rm FG} \sim t_{\Lambda}^{-2}  \sin^{-2}(\pi \tobs/t_{\rm f}), \ \  t_{\rm f}/2<\tc \ \ IV.
\end{equation}
The probability distribution is then
\begin{equation} 
\frac{d^3p_{\rm FG}}{d\log \tc~d\log \tl d \log \tobs}
\sim g \alpha \times \left\{\begin{array}{lll}
\vspace{.05in}\displaystyle{1 \over \tl^2 \tobs^2 }~, & \tobs<\tc
< \tl \ \ & I\\
\vspace{.05in}
\displaystyle{\tc \over  \tl^5 \sin^3(\tobs/\tl)}~,& \tc<\tobs<\tc^\prime  \ \ & II \\
\vspace{.05in}\displaystyle{1 \over
  \tl^2 (\tobs^\prime)^2} 
 ~, & \tc^\prime < \tobs \ \ &III \\
\vspace{.05in}\displaystyle{1 \over \tl^4 \sin^2(\pi \tobs/t_{\rm
    f})}~, & t_{\rm f}/2 < \tc \ \
& IV 
\end{array}\right.
\label{eq-pfgneq}   
\end{equation}

The probability force diagram is shown in Fig. \ref{fig-SFflow}.  At
fixed $\log \tobs$, the scale factor measure predicts that observers exist just before the crunch
($\log\tobs' \to \log {\tobs^{\rm min}}'$).  Recall that $\tobs^{\rm
  min}$ was introduced as a proxy for more detailed anthropic selection
effects.
The point is that the measure provides a pressure which favors
observers living as close as possible to the crunch. 

We can now find the probability distribution over $\tobs$. In the
previous sections, up to logarithmic corrections it did not matter whether  we optimized or
marginalized over $(\tl, \tc)$ because the distribution near the
maximum was exponential. Here, we will get different answers for the two procedures, so we choose to marginalize over $(\log t_{\Lambda},\log \tc)$, leaving
$\log \tobs$ to scan.  The resulting distribution is 
\begin{equation}
 \frac{dp}{d\log \tobs} \sim \tobs^{-3}\, \alpha(\log \tobs)~.
\label{eq-obssfn}
\end{equation}
There is no geometric pressure on $\log\tc$ in region III, where
Eq.~(\ref{eq-negsfprobcurv}) peaks, so the value of $\log \tc$ will
be determined by the prior distribution and anthropic selection.
Assuming that the prior favors small values of $\log \tc$, it seems
likely that expected value of $\log \tc$ is much less than $\log t_{\rm
  obs}$.  As in the apparent horizon and causal patch measures  for $\Lambda<0$, this complicates the
computation of $\alpha$.  However, the situation here is not
the same.  The difference is that here we have observers forming late
in the recollapse phase of a crunching universe, where the dominant
contribution to the energy density actually comes from matter.  The
fact that the universe is in a recollapse phase makes it very hard to
say what the form of $\alpha$ will be, whether or not there is an era of
curvature domination.

Regardless of the form of $\alpha$, the first factor in
Eq.~(\ref{eq-obssfn}) has a preference for $\log \tobs$ to be small.
If $\alpha$ it grows faster than $\tobs^3$, then it is favorable
for $\log \tobs$ to be large and $\log \tobs\to \log \tl^{\rm
  max}$~.  Otherwise, $\log \tobs \to \log \tobs^{\rm min}$, which means that some
anthropic boundary determines the expected value.

Now we will estimate the preference for negative values of $\Lambda$
over positive by integrating the distributions in Eqs.~(\ref{eq-obssfn}) and
(\ref{eq-obssfp}).  As mentioned above, to get agreement with the observed value of $\Lambda$ we need to assume $\alpha$ grows like a fourth power of $t_{\rm obs}$.  Then for both positive and negative $\Lambda$, the distribution is sharply peaked at $\tobs \sim \tl^{\rm max}$.   Then we find 
\begin{equation}
p_-/p_+ \sim \tl^{\rm max}~.
\end{equation}
So negative values of the cosmological constant are favored.

Finally, for $\Lambda<0$ it is worth noting the behavior of the
probability distribution over $\log \tobs$ for fixed $\log \tl$,
using for instance Eq.~(\ref{eq-negsfprob}) and neglecting for
simplicity the factor $\alpha$.  Depending on whether $t_{\rm
  obs}$ is larger or smaller than $t_{\rm f}/2$, $\log \tobs$ will be
driven either to $\log {\tobs^{\rm min}} '$ or to $\log \tobs^{\rm min}$.  The former case
is reproduced by our above procedure of fixing $\log \tobs$ and
letting $\log \tl$ vary.  The latter case is the time-reversed case
(and we know that the fat geodesic measure respects the time-reversal
symmetry of a crunching universe).  When both $\log \tl$ and $\log t_{\rm
  obs}$ are allowed to vary, we are driven to $\log t_{\Lambda}\sim \log t_{\rm
  obs}\sim \log \tobs^{\rm min}$ regardless of the order of scanning.

Recall that the fat geodesic cutoff is equivalent to the scale factor
measure in simple situations. However, our negative conclusions about
negative $\Lambda$ differ from the analyis of the scale factor measure
in \cite{DGSV08} which found no conflict with observation. There are
two reasons for this discrepancy. First, the fat geodesic
measure differs from the detailed prescription given in
\cite{DGSV08} in the recollapsing region. Second, the analysis of
\cite{DGSV08} made an unjustified approximation \cite{BouFre08b},
computing the scale factor time in the approximation of a homogeneous
FRW universe.
It remains to be seen if there is a precise definition of the scale
factor cutoff that will give the result computed in
\cite{DGSV08}. The fat geodesic is our best attempt to define a
simple measure in the spirit of \cite{DGSV08}.

\acknowledgments We thank Roni Harnik and Yasunori Nomura for interesting discussions.  This work was supported by the Berkeley
Center for Theoretical Physics, by the National Science Foundation
(award number 0855653), by fqxi grant RFP2-08-06, and by the US
Department of Energy under Contract DE-AC02-05CH11231. VR is supported
by an NSF graduate fellowship.

\bibliographystyle{utcaps}
\bibliography{all}
\end{document}